\documentclass[12pt]{article}
\usepackage{epsf}

\usepackage[nosort]{cite}
\usepackage{colortbl}

\usepackage{epsfig,graphics}
\usepackage {graphicx}
\usepackage {epsfig}
\usepackage {subfigure}
\usepackage {tabularx} 
\usepackage{rotate}	
\usepackage{slashed}

\usepackage{amsmath}
\usepackage{amsfonts}
\usepackage{amssymb}
\usepackage{graphicx}

\usepackage{fancybox,framed}
\usepackage[usenames,dviptsnames]{xcolor}

\usepackage[nottoc]{tocbibind}

\newcommand{\bmat}{\left(\begin{array}}
\newcommand{\emat}{\end{array}\right)}

\def\yzero{\smash{\hbox{$y\kern-4pt\raise1pt\hbox{${}^\circ$}$}}}

\def\beq{\begin{equation}}
\def\eeq{\end{equation}}
\def\beqa{\begin{eqnarray}}
\def\eeqa{\end{eqnarray}}

\def\-{\hphantom{-}}

\def\s2{\frac{1}{\sqrt2}}

\def\beq{\begin{equation}}
\def\eeq{\end{equation}}
\def\beqa{\begin{eqnarray}}
\def\eeqa{\end{eqnarray}}

\def\IF{\relax{\rm I\kern-.18em F}}
\def\II{\relax{\rm I\kern-.18em I}}
\def\IP{\relax{\rm I\kern-.18em P}}
\def\IC{\relax\hbox{\kern.25em$\inbar\kern-.3em{\rm C}$}}
\def\IR{\relax{\rm I\kern-.18em R}}

\def\Dsl{\,\raise.15ex\hbox{/}\mkern-13.5mu D} 
\def\IZ{Z\kern-.4em  Z}

\catcode`\@=11   
\newdimen\@rotdimen
\newbox\@rotbox  

\def\@vspec#1{\special{ps:#1}}
\def\@rotstart#1{\@vspec{gsave currentpoint currentpoint translate
   #1 neg exch neg exch translate}}
\def\@rotfinish{\@vspec{currentpoint grestore moveto}}
\def\@rotr#1{\@rotdimen=\ht#1\advance\@rotdimen by\dp#1%
   \hbox to\@rotdimen{\hskip\ht#1\vbox to\wd#1{\@rotstart{90 rotate}%
   \box#1\vss}\hss}\@rotfinish}
\def\@rotl#1{\@rotdimen=\ht#1\advance\@rotdimen by\dp#1%
   \hbox to\@rotdimen{\vbox to\wd#1{\vskip\wd#1\@rotstart{270 rotate}%
   \box#1\vss}\hss}\@rotfinish}%
\def\@rotu#1{\@rotdimen=\ht#1\advance\@rotdimen by\dp#1%
   \hbox to\wd#1{\hskip\wd#1\vbox to\@rotdimen{\vskip\@rotdimen
   \@rotstart{-1 dup scale}\box#1\vss}\hss}\@rotfinish}%
\def\@rotf#1{\hbox to\wd#1{\hskip\wd#1\@rotstart{-1 1 scale}%
   \box#1\hss}\@rotfinish}%
\def\rotate{\@ifnextchar[{\@rotate}{\@rotate[l]}}
\def\@rotate[#1]#2{\setbox\@rotbox=\hbox{#2}\@nameuse{@rot#1}\@rotbox}

\catcode`\@=12

\topmargin
-1.5cm
\textwidth
15.5cm
\textheight
23.5cm
\oddsidemargin
0.7cm
\evensidemargin
0.7cm

\definecolor{Gray}{gray}{0.95}
\definecolor{darkspringgreen}{rgb}{0.09, 0.45, 0.27}
\definecolor{darkseagreen}{rgb}{0.56, 0.74, 0.56}
\definecolor{darkmouthgreen}{rgb}{0.05, 0.5, 0.06}
\definecolor{darkcyan}{rgb}{0.0, 0.55, 0.55}
\definecolor{nicecolor}{rgb}{0.1, 0.3, 0.4}
\usepackage{color}
\definecolor{dark-gray}{gray}{0.20}
\definecolor{gray}{gray}{0.30}
\definecolor{light-gray}{gray}{0.80}
\definecolor{dark-red}{rgb}{0.7,0,0}
\definecolor{dark-green}{rgb}{0.1,0.4,0}
\definecolor{dark-blue}{rgb}{0.3,0.3,0.7}
\definecolor{light-blue}{rgb}{0.8,0.8,1}
\definecolor{swamp}{RGB}{240, 199, 197}

\definecolor{nicecolor}{rgb}{0.1, 0.3, 0.4}
\usepackage{jheppub} 
\hypersetup{
	colorlinks=true,
	linkcolor=dark-blue,
	citecolor=dark-red,
	urlcolor=nicecolor,
	linktoc=page,
	pageanchor=false
}

\usepackage[most]{tcolorbox}
\newtcolorbox{myconjecture}[1]{colback=green!5!white, colframe=green!45!black,fonttitle=\bfseries, title={#1}}
\begin{document}
\makeatletter
\@addtoreset{equation}{section}
\makeatother
\renewcommand{\theequation}{\thesection.\arabic{equation}}
\pagestyle{empty}
\vspace{-0.2cm}
\rightline{ IFT-UAM/CSIC-21-95}
\vspace{1.2cm}
\begin{center}


\LARGE{Swampland Constraints on Neutrino Masses\\
[13mm]}

  \large{E. Gonzalo$^*$, L.E. Ib\'a\~nez$^*$  and I. Valenzuela $^{**}$ \\[6mm]}
\small{
 *\  Departamento de F\'{\i}sica Te\'orica
and Instituto de F\'{\i}sica Te\'orica UAM/CSIC,\\[-0.3em]
Universidad Aut\'onoma de Madrid,
Cantoblanco, 28049 Madrid, Spain \\
**\ Jefferson Physical Laboratory, Harvard University\\[-0.3em]
Cambridge, MA 02138, USA }
\\[8mm]
\small{\bf Abstract} \\[6mm]
\end{center}
\begin{center}
\begin{minipage}[h]{15.22cm}
Compactifying the Standard Model (SM) on a circle may lead  to AdS 3D vacua, depending on the
 character (Majorana or Dirac) and the mass of the lightest neutrino. It has been shown that,  
 imposing the Ooguri-Vafa conjecture that no stable non-SUSY AdS vacua are consistent with Quantum
 Gravity, one can obtain conditions on the mass of the lightest neutrino. This result has the shortcoming that 
 it is in general  sensitive 
 to the UV structure of the theory.
  In the present paper we 
 show that two other independent swampland conditions may yield constraints very similar to those.
 These other two conditions come from the AdS swampland distance conjecture and 
 the dS conjecture as applied  to AdS vacua by Lust, Palti and Vafa. 
 Unlike the non-SUSY AdS constraints,  for these  conjectures the results require only  local IR information of the radion potential.
We consider  both the case of an explicit cosmological 4D constant and the alternative of
  a simple quintessence 4D  potential. 
Cosmological data in the next decade  may falsify the results,
  giving us information on the constraints of particle physics from Quantum Gravity.

\end{minipage}
\end{center}
\newpage
\setcounter{page}{1}
\pagestyle{plain}
\renewcommand{\thefootnote}{\arabic{footnote}}
\setcounter{footnote}{0}



\tableofcontents
\newpage
\section{Introduction}

Some of the most difficult problems in fundamental physics are associated to the understanding of mass scales. 
That is the case of the  Electro-Weak (EW) scale, $M_{\text{EW}}\simeq  10^2$ GeV,  whose value is sixteen orders of magnitude smaller than the scale 
of Quantum Gravity (QG), the (reduced) Planck scale $M_p=2.4\times 10^{18}$ GeV.
On the other hand, the accelerated expansion of the universe may be understood if there is a tiny cosmological constant 
vacuum energy with $V_0\simeq (2.4\times 10^{-3} \text{eV})^4$. This value is extremely small compared to both the EW and the Planck scales,
and this fact constitutes the cosmological constant problem. An alternative which has been proposed is the existence of a time dependent 
scalar potential asociated to a field, termed ``Quintessence" \cite{tsujikawa}. The observed acceleration of the universe would be associated to the
value of the scalar potential at present.  

There is also an interesting coincidence of fundamental mass scales, the scale of neutrino masses and the cosmological constant.
Neutrino oscillation \cite{Olive:2016xmw}  data give us  information about mass differences, with $\Delta m_\nu^2\simeq 10^{-3}-10^{-5}$ $\text{eV}^2$ and cosmological
limits see e.g. \cite{GAMBIT,olga} tell us that $\sum m_{\nu_i}\lesssim 0.1$ eV\footnote{Note however that the lightest neutrino being massless is not excluded experimentally.}. This range of masses is very close to the c.c. scale mentioned above. 
The origin of both quantities is very different, with the neutrino masses coming from EW symmetry breaking and (possibly) the breaking of
lepton number. On the other hand all dynamical potentials in the theory from all sectors contribute to the value of the cosmological constant.

Recently it has been pointed out that consistency with Quantum Gravity could be at the root of the numerical coincidence between the scale of neutrino
masses and the cosmological constant  \cite{Ooguri:2016pdq,IMV1,HS,tom} see also \cite{Lust}.
The idea is related to the ``Swampland Program" \cite{swampland} whose objective is to identify which classes of
effective field theories may have an UV completion consistent with Quantum Gravity (see \cite{vafafederico,review,irenereview,alvaroreview} for reviews). 
Essentially, the prediction is that 1) the lightest neutrino cannot be Majorana and 2) the Dirac (or pseudo-Dirac) mass of the lightest neutrino must verify
\beq
m_{\nu_1}\ \lesssim \ \Lambda_4^{1/4} \ ,
\label{guay}
\eeq
we will be more precise in the main text.
The key idea  in the Swampland Program is that 
most of the effective  field theories which look consistent from the low energy physics point of view (i.e. anomaly-free, unitary...) cannot
be embedded into a consistent theory of quantum gravity. Some examples of this fact are surprising: a theory with free photons coupled to gravity 
is not consistent \footnote{For it to be consistent charged states must be added with at least one of them verifying $m\leq \sqrt{2}qM_p$    \cite{WGC} .}.
Many properties of the Swampland are formulated in terms of conjectures, with diverse degree of rigour. The above constraint on the lightest neutrino mass
is based on the conjecture \cite{Ooguri:2016pdq}:
\begin{itemize}
\item 
{\it AdS non-SUSY conjecture}. There are no stable non-SUSY AdS vacua consistent with 
quantum gravity.
\end{itemize}
The connection to  neutrino masses is as follows. If we compactify the SM on a circle, the radion scalar gets a potential in 3D with two terms,
a tree level one from the 4D c.c. and a one-loop Casimir potential, which has a neutrino mass dependent piece.
  It was shown in \cite{ArkaniHamed:2007gg} that  AdS vacua may form in 3D, depending on the mass of the lightest neutrino. Such  vacua, if stable, 
  would thus violate the above conjecture, in contradiction with our 4D SM vacuum being consistent with QG. However one finds that these AdS vacua do not form if the lightest neutrino is sufficiently light, 
  i.e. $m_{\nu_1}\ \lesssim \ \Lambda^{1/4} $.  Hence, this bound is a sufficient condition to avoid trouble with the conjecture at low energies. It has also been argued recently that the presence of light fermions in dS backgrounds obeying that type
  of conditions could be a general fact in Quantum Gravity \cite{lightfermion}. 

While  this connection is quite elegant and intriguing,  the conjecture requires that the AdS vacua in 3D should be stable in order for the inconsistency to arise. 
One may think of possible non-perturbative instabilities like the creation of ``bubbles of nothing" \cite{bubbles}. These are in fact not possible in this setting because of the
periodic boundary conditions assigned to the SM fermions, unless one adds somehow some UV spin defect that trivializes the corresponding cobordism group \cite{McNamara:2019rup}.  The presence of EM Wilson lines on the circle may also lead to instabilities \cite{HS}  which may be 
however avoided in related settings using orbifold projections \cite{2toro}. Still, it is not easy to guarantee the complete stability of the vacua without having a more
complete information about the UV physics.

In the present paper we show that other totally different  Swampland conjectures give rise to analogous constraints on neutrino masses,
reinforcing the qualitative statement in Eq. (\ref{guay}). We also  find that this is true also in the case in which one has a simple Quintessence potential
instead of a cosmological constant. 
The first conjecture we use is the \cite{lpv}
\begin{itemize}
\item {\it AdS Swampland distance conjecture}. 
 Consider quantum gravity on a
AdS space with cosmological constant $\Lambda_{0}=V_{0}$. 
If there is a family of minima such that $\Lambda_{0}\rightarrow0$,
then an infinite tower of states  should appear with mass scale $m$ behaving  (in Planck units) as 
\begin{equation}    
m\sim\left|\Lambda_{0}\right|^{\alpha}=\left|V_{0}\right|^{\alpha} \ ,
\end{equation}    
where $\alpha$ is a positive order-one number.
\end{itemize}
This conjecture has been shown to be correct in many string theory vacua up to now, with no fully controlled top-down counterexamples.
We will apply this conjecture to the AdS 3D vacua of the SM mentioned above. An important difference compared to the 
AdS non-SUSY conjecture applied in previous work is that the conjecture does not require the minima to be stable.
Any local or metastable minima must verify the conjecture, and hence the results are independent from UV physics. We presented some of the implications arising from applying the conjecture to compactifications of general D-dim vacua in \cite{lightfermion}. In this paper, among other things, we will carry out a much more detailed analysis for the case of the SM.
The main assumption will be that the neutrino masses scan in the landscape of string theory solutions. 
If one keeps fixed the c.c. to its experimental value, one finds that the above conjecture requires that the lightest neutrino is
not Majorana and has a Dirac mass $m_{\nu_1}\leq 7.7$ meV (for normal hierarchy NH), very similar to the results mentioned above.
We also consider the case in which both neutrino masses and
cosmological constant scan in a landscape, i.e., $m_{\nu_1}\rightarrow \lambda m_{\nu_1}$,$\Lambda\rightarrow \lambda^\alpha \Lambda$.
We find that for $\alpha<4$ one obtains the same bounds on the lightest neutrino. The opposite bound is obtained for scannings with $\alpha>4$.

 To obtain the above constraints, we assumed that the we live in a (meta)stable dS vacuum with a 4D cosmological constant  that we use as an input.
As we mentioned, it has been suggested that the accelerated expansion of the universe may be rather due to the runaway 
scalar potential of a Quintessence field. 
In particular it has been put forward a \cite{dS1,dS3,Krishnan}
\begin{itemize}
\item {\it dS  Swampland conjecture}. In a consistent theory of quantum gravity 
the scalar potential must verify
\begin{equation}
 {\left|\nabla V\right|}   \ \geq \ c {V}\ 
\label{dsone}
\end{equation}
with $c$ a positive constant of order one. 
\end{itemize}
Such a constraint is expected to be valid at the asymptotic regimes of the field space, although not necessarily deeper in the bulk. See \cite{dS3,TCC,Dvali:2014gua,Dvali:2017eba} for refinements and related proposals. In particular, it implies that dS vacua are problematic (at least, for large field ranges) in a theory of QG so that something like Quintessence may be forced eventually upon us. 
Thus, an obvious question is whether one may obtain again constraints on neutrino masses if assuming a quintessence-like scenario instead of a 4D dS vacuum.
When compactifying the SM on a circle we have then two scalars in 3D, the radion and the Quintessence one, and it turns out one does not find
local AdS minima for any value of the neutrino masses. Thus we cannot apply the above two Swampland conjectures to constraint 
the lightest neutrino mass. However, non-trivial constraints appear from applying a refinement of the above dS conjecture in which  one has $\left|V\right|$ instead of $V$ in \eqref{dsone}. This refinement was proposed in \cite{lpv} by investigating the  generalisation to AdS vacua.
%
Interestingly, imposing such a constraint to a canonical, single-field Quintessence 
model added to the SM we again find constraints on the value of the lightest neutrino mass. The precise bound depends on the
constant $c$ and the corresponding parameter $c_V$ which controls the Quintessence potential. Setting those parameters to natural values,
consistent with cosmological constraints, one gets neutrino mass bounds on the same ballpark, $m_{\nu_1}\lesssim 8$ meV.

The structure of this paper is as follows. In the next section we review previous results which find constraints on the lightest
neutrino mass from the non-SUSY AdS conjecture. This also allows us to describe the compactification of the SM on a circle and fix the notation.
In section 3 we make use of the AdS distance conjecture to constraint the  lightest neutrino mass. We consider several choices of landscape parameter scannings
both for neutrino masses and the 4D cosmological constant. In section 4 we consider  the case in which the 4D cosmological constant is replaced by
a Quintessence potential. We study how applying the dS Swampland conjecture, as extended to AdS vacua, gives again bounds on the lightest neutrino mass.
We leave the 5-th section for some general conclusions.

\section{The SM on the circle and the non-SUSY AdS constraint}
\label{introneu}

In this section, we review the compactification of the SM on a circle \cite{ArkaniHamed:2007gg} and previous results \cite{IMV1,IMV2,HS,2toro,Gonzalo} obtained from applying the Non-SUSY AdS conjecture \cite{Ooguri:2016pdq}. The rationale behind compactifying the SM to get constraints on the 4D field spectra from imposing the Swampland conjectures will be present throughout the entire paper 
and goes as follows. By background independence of quantum gravity, we expect that any compactification of a consistent theory should also be consistent\footnote{Even though sometimes there might be obstructions coming from fermionic or other topological structures, none of them are known to apply to the case of circle compactifications, which is all we need for the purposes of this paper.}. This opens up an interesting avenue to discern whether a given EFT is consistent with quantum gravity: by checking whether it remains consistent upon compactification. Therefore, if we find that the SM compactified on a circle  does not satisfy the Swampland conjectures, it implies that the SM itself living in four dimensions was already inconsistent with quantum gravity. Since this outcome will depend on the masses of the light field spectra in our universe, we can use this logic to provide bounds on neutrino masses that ensure that compactifications of the SM are consistent with the Swampland conjectures.

\subsection{The SM on a circle}
 We begin by considering the $4d$ low-energy effective action of the SM coupled to Einstein gravity
 \cite{ArkaniHamed:2007gg,IMV1,2toro,HS}
\begin{align}
S & =\frac{1}{2}\int d^{4}x\sqrt{-g_{4}}\biggl\{(M_{p}^{4d})^{2}(\mathcal{R}^{4d}-\Lambda^{4d})-\frac{1}{4}F_{\mu\nu}F^{\mu\nu}\\
& +\sum_{i=1}^3\bar{\nu_{i}}\left(i\tilde{\gamma}^{\mu}D_{\mu}-m_{\nu_{i}}\right)\nu_{i}\biggr\}+\dots
\end{align}
where $F_{\mu\nu}$ is the field strength of the photon and $\nu_i$ are the neutrinos. The rest of the SM particles can be safely ignored for the purposes of this paper, since only the lightest SM fields will play a role in the computation of the Casimir energy potential upon compactification.

In the SM (active, left-handed) neutrinos are massless. However, neutrino oscillation experiments \cite{neutrinooscila1,neutrinooscila2,neutrinooscila3} determined that at least two neutrinos are massive in order to account for the following measured neutrino mass differences \cite{Olive:2016xmw}
\begin{eqnarray}
\Delta m_{21}^2 = (7.53\pm 0.18)\times 10^{-5}\,\, {\rm eV}^2,\\
\Delta m_{32}^2 = (2.44\, \pm 0.06)\times 10^{-3}\,\, {\rm eV}^2 \,\,{\rm (NH)},\\
\Delta m_{32}^2 =(2.51\pm 0.06)\times 10^{-3}\,\, {\rm eV}^2 \,\,{\rm (IH)}.
\end{eqnarray}
In this work, we consider constraints on the simplest BSM theories where the SM is minimally extended with three sterile right-handed Weyl neutrinos in order to account for neutrino masses. We will consider the possibility of having both Dirac and Majorana masses. Typically, Majorana spinors are considered in the context of the See-Saw mechanism, where right-handed neutrinos are very massive and that forces active neutrinos to be very light. This implies that the number of light degrees of freedom is different if considering Dirac masses or this See-Saw mechanism. In this paper, we will refer to Majorana neutrinos when considering only 6 light degrees of freedom (2 per neutrino), while Dirac neutrinos will refer to the case of having 12 light fermionic degrees of freedom. This different counting will play an essential role in our analysis.


Let us now compactify the SM in a circle such that one of the spatial dimensions only takes values on $x_{3}\in\left(0,2\pi\right)$ and study the dimensionally reduced 3d action. For simplicity, we will consider a circle compactification without Wilson lines. The metric  is given by \begin{equation}
g_{\mu\nu}=\left[\begin{array}{cc}
\frac{r^{2}}{R^{2}}g_{ij} & 0\\
0 & R^{2}
\end{array}\right],
\label{metric}
\end{equation}
where we have ignored the graviphoton since it will not play any role in our analysis, and the scalar field $R$ parameterizes the radius of the circle.
This gives the following one-loop corrected quantum effective action in 3d:

\begin{equation}
\Gamma  =\int d^{3}x\sqrt{-g_{3}}\biggl\{ M_{p}^{3d} \frac{1}{2}\mathcal{R}^{3d}+M_{p}^{3d} \frac{\partial^{i}R\partial_{i}R}{R^{2}}-V(R) \biggr\}
\label{reduction}
\end{equation}
\beq
V(R)=V_{\rm tree}(R)\ +\ V_{1L}(R)\ , \quad V_{\rm tree}(R)=2\pi r\left(\frac{r}{R}\right)^{2}\rho_{\Lambda^{4d}}
\eeq
\begin{equation}
V_{\text{1L}}(R)=-\frac{n_{b}r^{3}}{720\pi R^{6}}+\sum_{i=1}^{3}n_{\nu}\frac{r^{3}m_{\nu_{i}}^{2}}{4\pi^{3}R^{4}}\sum_{n=1}^{\infty}\frac{K_{2}(2\pi nRm_{\nu_{i}})}{n^{2}}\cos n\theta
\label{potentialuno}
\end{equation}
Here, $n_{b}$ is number of massless bosonic degrees of freedom (4 in the SM coming from the photon and the graviton), $n_{\nu}$ is the number of degrees of freedom of each fermion (4 if neutrinos are Dirac and 2 if they are Majorana), $M_{p}^{3d}=2\pi r\left(M_{p}^{4d}\right)^{2}$ and $\rho_{\Lambda^{4d}}=\left(M_{p}^{4d}\right)^{2}\Lambda^{4d}=2.6\times 10^{-47} \text{GeV}^4$ . We will consider only the case where neutrinos have periodic boundary conditions on the circle, so $\theta=0$.

\begin{figure}
	\centering{}\includegraphics[scale=0.40]{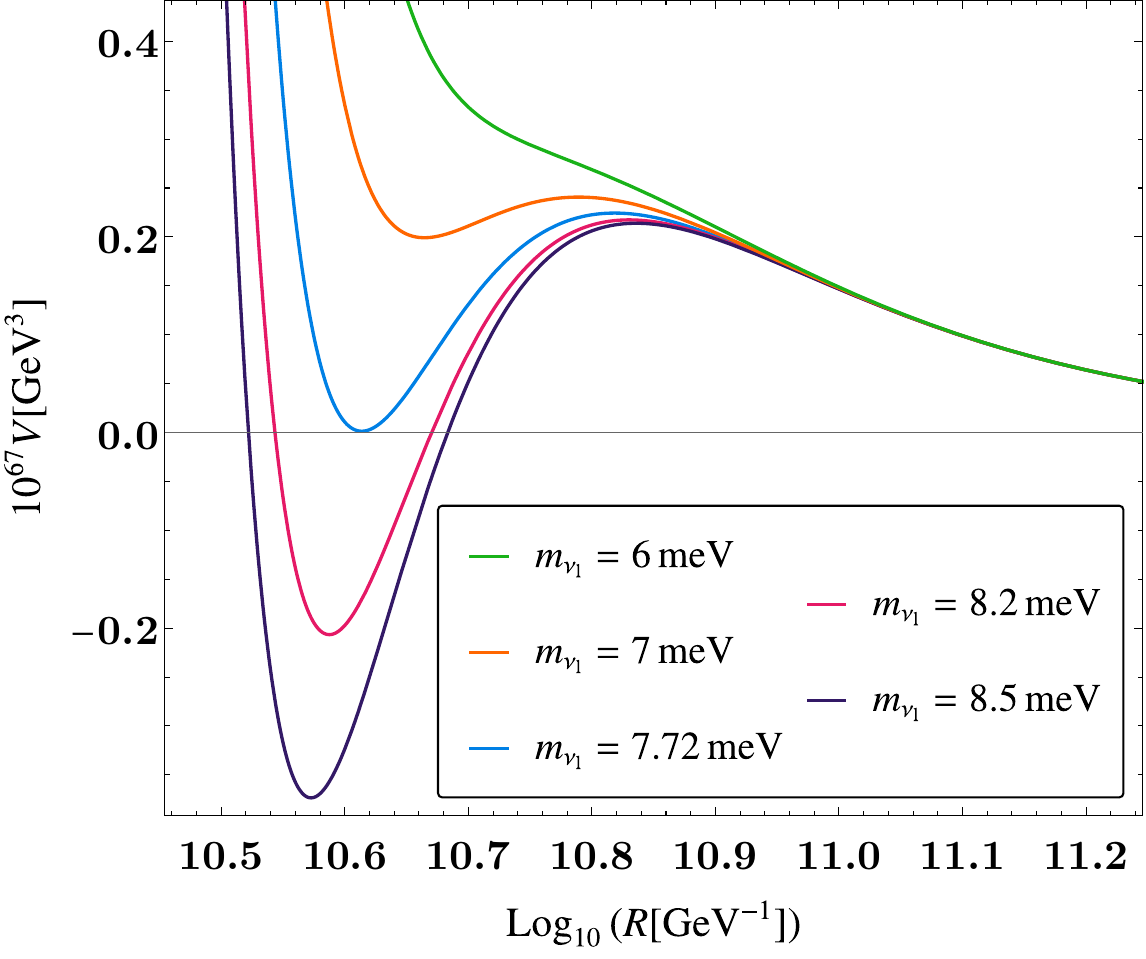} \
	\caption{The radion scalar potential for different values of the lightest Dirac neutrino mass.}
	\label{adcpotential}
\end{figure}

In addition to the tree level contribution to the potential $V_{\rm tree}$  coming from the 4d cosmological constant, there is a one loop correction $V_{1L}$ coming from integrating out all the Kaluza-Klein copies of the 4d SM particles, yielding the infinite sum in $n$ above. This is known as the Casimir energy potential and it becomes important only for 4d particles with a mass smaller than $1/R$.


The goal of this paper is to determine the constraints on the neutrino masses that guarantee that this 3d potential satisfies the Swampland conjectures described in the introduction. To illustrate this, we plot in Fig. \ref{adcpotential} the potential as a function of the radius for different values of the neutrino masses for the case of Dirac neutrinos with NH. Notice that the potential will approach zero from above as $R\rightarrow \infty$ since the tree level contribution from a positive 4d cosmological constant dominates for large radius. As we decrease the radius, the negative contribution from the massless photon and graviton dominates, so that the potential develops a maximum around $R\sim \Lambda_4^{1/4}$ and starts decreasing. Since fermions contribute positively to $V_{1L}$, they will change the tendency of the potential at some point and may generate a minimum in AdS, Minkowski or dS, depending on how their mass compares with $\Lambda_4^{1/4}$. Let us note that the above plot is constructed by preserving the experimental neutrino mass differences and changing the mass of the lightest neutrino. Under these conditions, an $AdS_{3}$ minima, i.e. $V(R_{0})=V_{0}<0$,  will form 
if the mass of the lightest neutrino is large enough compared with $\Lambda_4$, namely $m_{\nu_1}\leq 7.7\times 10^{ -3}$ eV  when replacing the value of  $\Lambda_4$ in our universe \cite{IMV1}. 
As we will see momentarily, in the case of Majorana neutrinos the formation of such  AdS vacua turns out to be unavoidable, for any value of the lightest neutrino mass.

\subsection{Review of constraints from the Non-SUSY AdS conjecture}

 According to the Non Susy AdS Instability conjecture that we described in the Introduction, stable non-SUSY AdS vacua are in the Swampland. Hence, if the non-supersymmetric $AdS_3$ vacua arising from compactifying the SM were stable, the theory would be inconsistent with quantum gravity.  This can be avoided by imposing constraints on the 4d light spectra to prevent these vacua to form. Several interesting conclusions have been obtained in the literature by imposing the absence of AdS minima, as we will review in the following.

First of all, neutrinos cannot be purely Majorana but must have a comparable Dirac mass, since otherwise an AdS vacuum will necessarily form. Thus, the simplest See-Saw mechanism for Majorana neutrinos seems inconsistent with QG \cite{Ooguri:2016pdq,IMV1}.
Furthermore, the absence of AdS minima imposes an upper bound on the Dirac mass of the lightest neutrino such that only metastable dS vacuum or runaways are generated. For the simple circle compactification of Fig. \ref{adcpotential} we can see that the the bound in NH is $m_{\nu_1}\leq 7.7\times 10^{ -3}$ eV .  For the case of inverted
hierarchy (IH) one gets a slightly stronger constraint $m_{\nu_1}\leq 2.5 \times 10^{ -3}$ eV \cite{IMV1}.
Similar bounds follow from toroidal and orbifold compactifications to two dimensions \cite{2toro}, where the Wilson lines are projected out so that one can avoid possible non-pertubative instabilities pointed out in \cite{HS}. 
This yields $m_{\nu_1}\leq 4\times 10^{ -3}$ eV (normal hierarchy) or $m_{\nu_3}\leq 1\times 10^{ -3}$ eV (inverted hierarchy) \cite{IMV1} for a toroidal compactification. These results can also be interpreted as lower bounds for the cosmological constant $\Lambda_4$ in terms of the mass of the lightest neutrino, $\Lambda_4\gtrsim m_{\nu_1}^ 4$ \cite{IMV1}, relating two a priori disconnected scales in Cosmology and Particle Physics, and forbidding Minkowski vacua, unless the lightest neutrino was massless.
In fact, for fixed  neutrino Yukawa couplings, they translate to a an upper bound on the Electro-Weak  scale in terms of the cosmological constant \cite{IMV1,IMV2}, bringing new insights into the EW hierarchy problem.
Finally, these ideas can also be used to show that the EW scale must be larger than the QCD scale \cite{Gonzalo}.

The big caveat of all these results is the assumption about vacuum stability. Recall that for these AdS vacua to be inconsistent with quantum gravity they must stable. However, even if they seem stable from the IR perspective, there could be hidden UV non-perturbative stabilities that cannot be read from the IR radion potential. If that was the case, then the constraints on neutrino masses would not apply. Unfortunately, this is something that cannot be determined without further UV information. In fact, even if our vacuum in four dimensions is unstable, the instability will be transferred to three dimensions only if the bubble radius is smaller than the $AdS_3$ scale, so it becomes a model-dependent question, sensitive to the specific properties of the instability. To overcome this issue, in this paper we are going to study the constraints on the SM coming from other Swampland conjectures that do no require this assumption about vacuum instability. Interestingly, we will obtain similar bounds on neutrino masses, reinforcing the idea that the coincidence between the scale of the cosmological constant and neutrino masses has a deeper quantum gravity origin.

\section{The AdS distance conjecture and neutrino masses}

In this section, we are going to explore the constraints coming from the AdS distance conjecture (AdC) when applied to the 3D
circle compactification of the SM. This conjecture can be applied whenever there is a family of vacua with different values of the vacuum energy  approaching the flat space limit $\Lambda\rightarrow 0$. Hence, we are going to assume that the SM is part of a landscape of vacua that differ by the value of the SM particles masses. This generates different values of the 3D cosmological constant. By scanning on the neutrino masses, we will then scan different 3d vacua that will get constrained by the AdC. The results will therefore depend on the specific scanning chosen within the space of parameters of the SM. This brings a certain level of ambiguity, since we do not really know what scans produce a family of vacua consistent with quantum gravity, so we will consider several possibilities in this section. The advantage, though, is that the results will be independent of the stability of the vacuum, unlike those coming from the Non-SUSY AdS conjecture.  Both conjectures, therefore, complement each other and give rise to similar results.

As we said, we are making here the  assumption that the string theory underlying the SM is such that neutrino masses scan in a landscape of possible values, so let us motivate this assumption before presenting the results. 
From the point of view of string theory such a scanning would have its origin in the huge number of possible  fluxes in generic compactifications, each of them leading to
different values of parameters in the EFT.
Such  scanning  is also  in particular required  if the EW hierarchy problem has an anthropic explanation. It is well known that  e.g. the vev of the Higgs 
cannot  be too far from its observed value, otherwise complex and stable nuclei would not form 
\cite{Damour:2007uv,Donoghue:2009me,Hall:2014dfa,Meissner:2014pma,Donoghue:2016tjk}.
Moreover, at least the masses of the first generation
quarks and leptons need to scan in an anthropic setting, since e.g. the strength of nuclear forces and the stability of the proton sensitively depend on them
\cite{Damour:2007uv,Donoghue:2009me,Hall:2014dfa,Meissner:2014pma}.
 Independently of any anthropic consideration, the fact that stability of matter depends on very particular choices of parameters suggests that there  is a plethora of vacua very similar to our own, a landscape of nearby vacua. Otherwise one would have to accept that the  observed values of the lightest
fermion masses and the Higgs vev are the only possible output of the fundamental theory, which would be, in our opinion,  quite unlikely.
In any event, we will assume in what follows that neutrino masses  scan in the fundamental theory
\footnote{ For recent references to models  in which the Higgs vev scan in a landscape see \cite{Herraez:2016dxn,Giudice:2019iwl,Kaloper:2019xfj}}.

First of all, we will study the simplest case where the string theory underlying the SM is such that the cosmological constant  $\Lambda_4$ 
remains unaffected by the scanning on neutrino masses. This will be relaxed in section 3.3. In this case, we have a three-dimensional parameter space given by the three neutrino masses. As explained above, the rest of SM particles play no role on the generation of minima. We are going to focus on those scans that seem better justified. For instance, we will extensively study
an {\it homogeneous} scanning in subsection \ref{sec:homogeneous} where  all neutrino families scale in the same way under a rescaling by a
parameter $\lambda$,  so that $m_i(\lambda)= \lambda  m_i^\text{exp}$, where $m_i^{\text{exp}}$ are the physical values of the neutrino masses in our universe. There are therefore two free parameters, $\lambda$ and $m_1^{\rm exp}$, since the mass of the other two neutrinos are fixed by neutrino mass oscillations in terms of the first one. One can then think of $\lambda$ as the Higgs vev or some homogeneous scanning on the  neutrino Yukawas. 
Next, in section \ref{sec:nonhomogen} we will consider scannings where only the mass of the lightest neutrino changes, so that they correspond to varying the first Yukawa coupling. 
Finally, in section \ref{sec:lambdachange} we will consider scannings where the 4d cosmological constant also changes.

\subsection{Homogeneous scanning}
\label{sec:homogeneous}
 Neutrino masses will vary with the scanning parameter $\lambda$ as
\begin{align}
m_{\nu_{1}}(\lambda) & =\lambda m_{\nu_{1}}^{\text{exp}}\nonumber \\
m_{\nu_{2}}(\lambda) & =\lambda\sqrt{(m_{\nu_{1}}^{\text{exp}})^{2}+\Delta m_{21}^{2}}\nonumber \\
m_{\nu_{3}}(\lambda) & =\lambda\sqrt{(m_{\nu_{1}}^{\text{exp}})^{2}+\Delta m_{21}^{2}+\Delta m_{32}^{2}}
\end{align}
In the case of Dirac masses, one has $m_i^\text{exp}=Y_i|H_0|$ with $|H_0|$ the physical
Higgs vev.  
Thus, the scanning in $\lambda$ may be interpreted as a scanning on the Higgs vev, $m_i(\lambda)= Y_i |H_0^\lambda|$,
with $H_0^\lambda\equiv \lambda H_0$. In the case of Majorana masses the same applies with $H_0$ replaced (in e.g. the seesaw
mechanism) by $|H_0|^2/|N_0|$, with $N_0$ a large (lepton number violating) scalar vev. An homogeneous 
scanning may also be reinterpreted as a uniform scanning of  the Yukawa couplings, i.e.,  $Y_i(\lambda)=\lambda Y_i^0$. 

A scanning parameter $\lambda$ does not need to vary  from zero to infinity, but may be limited to some range of values.
Thus e.g., in flux string
compactifications the scanning may arise from a multitude of values for discrete fluxes, but the ranges of fluxes are 
typically limited by constraints like tadpole cancellation or absence of Freed-Witten anomalies. Hence we will 
allow for limited ranges for the scanning parameter $\lambda_{\text{min}}\leq \lambda \leq \lambda_{\text{max}}$. Note that any such range
should contain the point $\lambda=1$ corresponding to the value of neutrino masses realized in nature. We will also assume that the point $\lambda\simeq 0$ is also part of the landscape, corresponding to a vacuum with approximately massless neutrinos and/or restored EW symmetry breaking.\footnote{Actually, it was argued in \cite{Gonzalo} that a zero Higgs vev is not allowed in the landscape since it gives rise to additional $AdS$ vacua that would be inconsistent with the Non-SUSY AdS conjecture.  A non-zero Higgs vev was also recently motivated in \cite{festinanew} by another swampland conjecture named Festina Lente \cite{festina}. However, for our purposes, it is enough to scan until a very small (non-zero) $\lambda$, say e.g. $\lambda\sim 10^{-4}$ , still consistent with \cite{Gonzalo,festinanew}.} We will therefore consider the following range for the scanning parameter,
 $0\lesssim \lambda\leq \lambda_\text{max}$, with $\lambda_\text{max}\geq 1$.

As explained in the previous section, different neutrino masses yield different 3d radion potentials upon circle compactification of the SM. Hence, this homogeneous scanning translates into a scanning of
a family of 3d vacua with different vacuum energy. This family of vacua will be subject to constraints from the 
above discussed AdS distance conjecture when the scanning drives $V_0\rightarrow 0$. 
These constraints depend on the
number of neutrino degrees of freedom, as we now discuss.

Consider first the case in which neutrinos are Majorana, with only two degrees of freedom per generation. 
In this case, for very small $\lambda$   we would have altogether 6 light fermionic degrees 
of freedom from the three neutrino generations,  which win over the 4 bosonic degrees of freedom of the photon and graviton. Hence the potential is 
a runaway for small $\lambda$. However, as $\lambda$ (e.g. the Higgs vev) increases, the neutrino masses start to split and the lightest, if Majorana,
has only 2 degrees of freedom. That is insufficient to stop the potential from becoming negative as $\lambda$ increases.
Eventually, however, for small enough radius the two heaviest neutrinos contribute and the potential becomes again positive. 
Thus we smoothly go from positive runaway  to negative minima at finite value of the radius.  
 Scanning back, we see that we can go smoothly from an AdS vacuum to Minkowski.
This is forbidden by the AdS conjecture, unless some infinite tower  of states appears precisely when the potential hits $V_0=0$.
But the only tower that we have in the system is the KK tower which only becomes light when $R\rightarrow \infty$ and not at   finite $R$. 
Hence Majorana neutrino masses would be ruled out, under the assumption that this is a consistent scanning in quantum gravity.

\begin{figure}
	\centering{}\includegraphics[scale=0.3803]{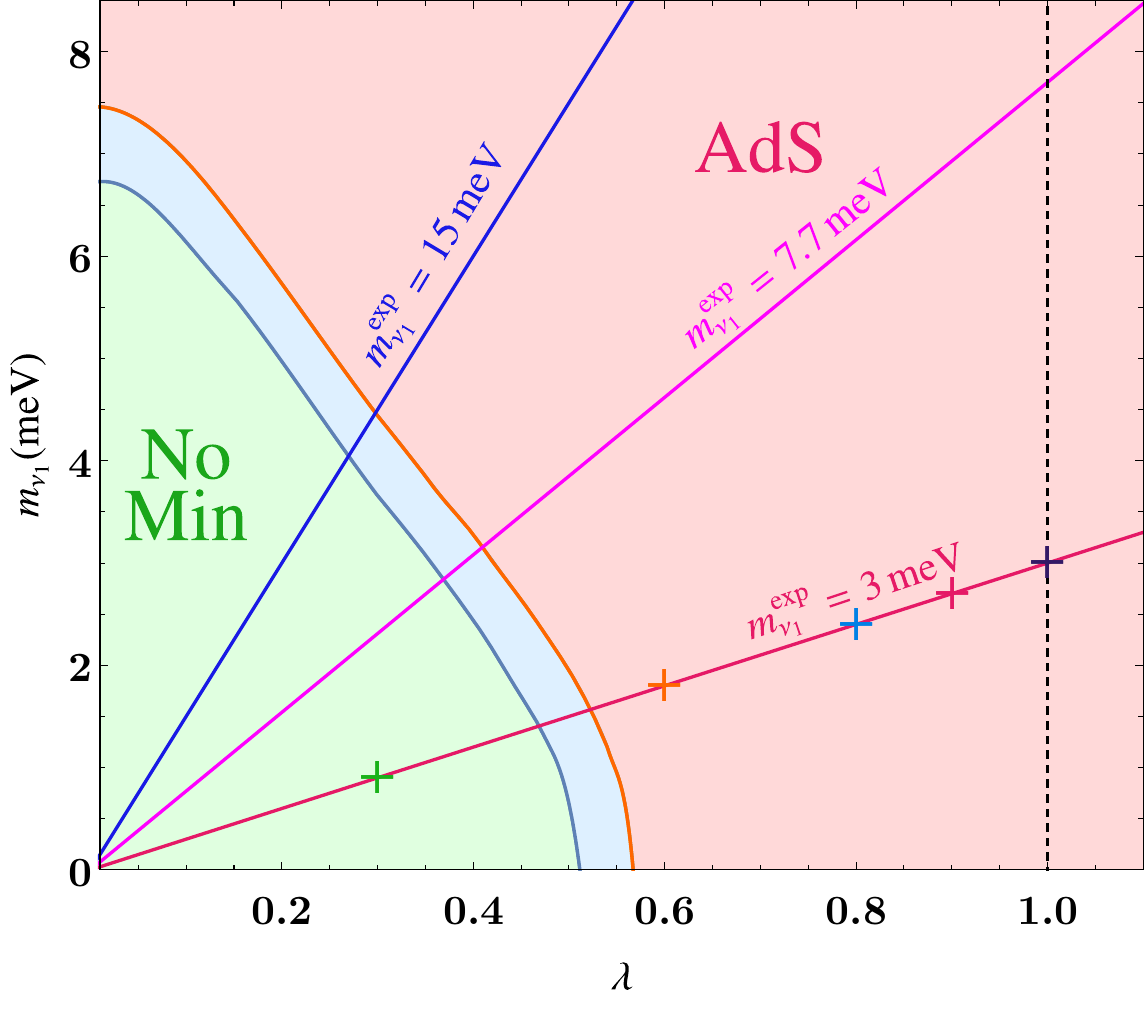} 
	\includegraphics[scale=0.41]{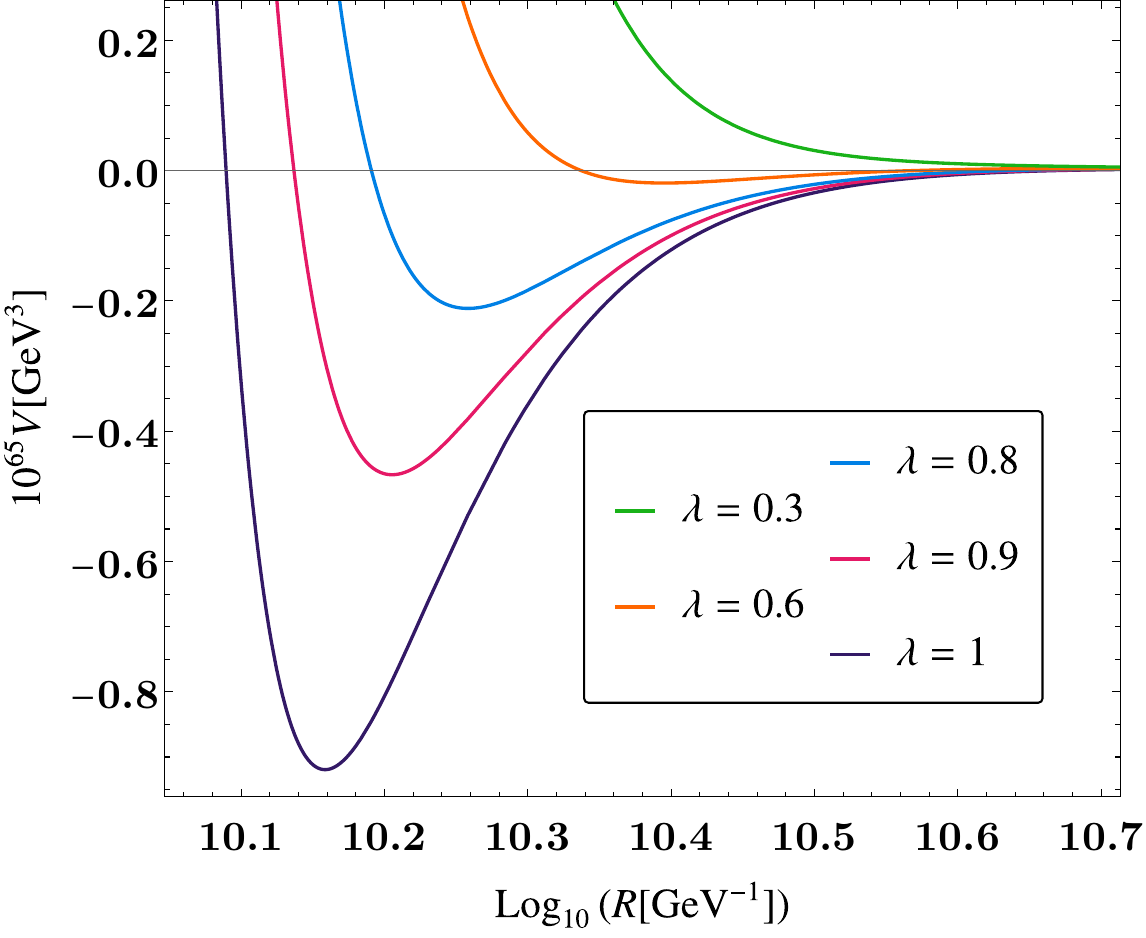}
	\caption{Left: Lightest neutrino mass versus the scanning parameter $\lambda$ for Majorana neutrinos.
		The green area corresponds to runaway dS vacua and the pink one to AdS vacua, with a narrow region of dS vacua in between.
		Straight lines correspond to  homogeneous scans for different physical neutrino masses. The lines 
		necessarily cross from the AdS to the runaway area, violating the AdS distance conjecture. Right: In this figure we show how the potential changes along the scanning for the particular value  $ m_{\nu}^{\text{exp}} = 3$ meV. Coloured crosses in the left hand side correspond to the coloured curves in the plot on the right hand side.}
	\label{majorana_cero}
\end{figure}

The results for the Majorana case are illustrated in Fig. \ref{majorana_cero}. In the left-hand side we plot the mass of the lightest neutrino versus the scanning parameter lambda. Each of the straight lines correspond to an homogeneous scanning of neutrino masses for different values of the lightest neutrino Yukawa coupling, corresponding to different masses for the lightest neutrino for $\lambda=1$, as shown in the plot.   Note that the lines go continuously from a positive runaway (green region) , to  dS  (blue region) and to AdS (light red region)
as we decrease $\lambda$, 
indicating a violation of the AdS conjecture.
 This happens for any value of the first generation Yukawa coupling, and cannot be avoided by 
choosing  a smaller range $\lambda\lesssim 0.5$ since, as we said, any scanning should contain the value $\lambda=1$ which corresponds to the experimental result. In the right-hand side we plot the potential for the particular case where the experimental mass of the neutrinos is 3 meV. Each coloured cross in the left plot corresponds to a curve of the same colour in the right-hand side plot. One can see, again, that we can continuously cross the flat space limit and go from dS to AdS vacua by varying $\lambda$.  This is inconsistent with the AdS Distance conjecture, since there is no tower of states becoming massless when crossing Minkowski as explained above.

\begin{figure}
	\centering{}\includegraphics[scale=0.38]{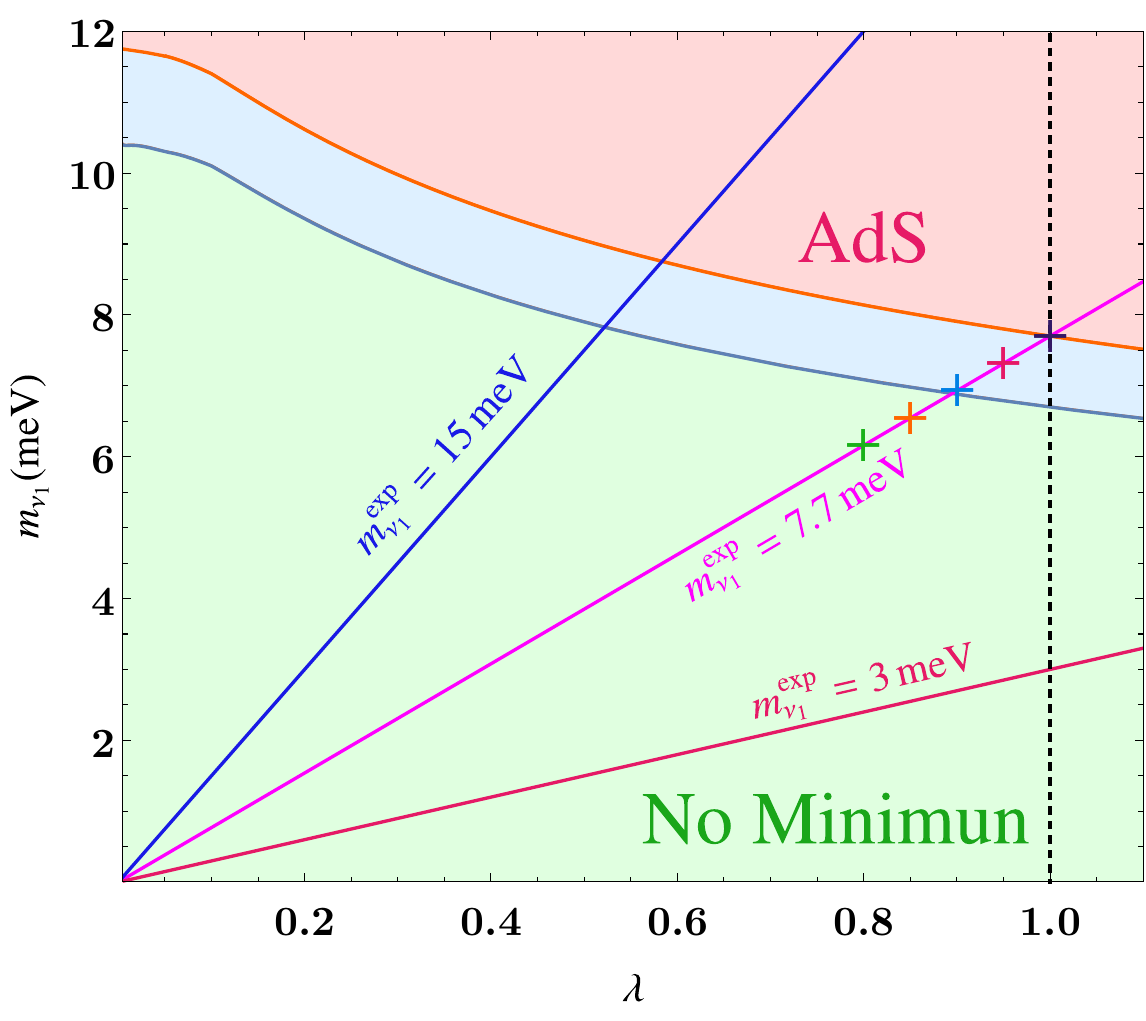} 
	\includegraphics[scale=0.3905]{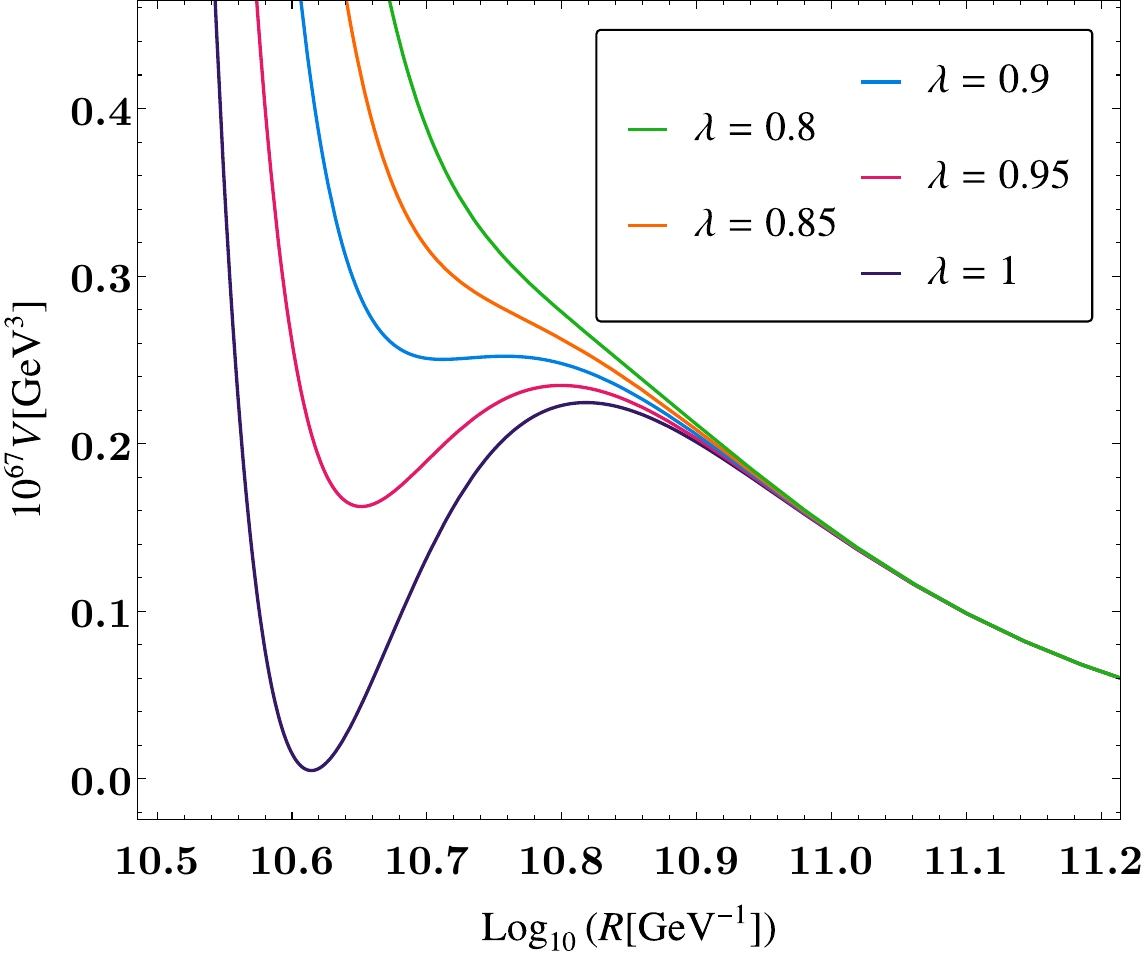}
	\caption{Left: Lightest neutrino mass versus the scanning parameter $\lambda$ for Dirac neutrinos. 
		The green area corresponds to runaway dS vacua and the pink one to AdS vacua, with a narrow region of dS vacua in between.
		Straight lines correspond to  homogeneous scans for different physical neutrino masses. The lines 
		do not cross from the AdS to the runaway area as long as the lightest neutrino mass is $ m_{\nu}^{\text{exp}} \leq 7.7$ meV. Right: In this figure we show how the potential changes along the scanning for the particular value  $ m_{\nu}^{\text{exp}} = 7.7$ meV. Coloured crosses in the left hand side correspond to the coloured curves in the plot on the right hand side.}
	\label{dirac_cero}
\end{figure}

Consider now the case in which neutrinos are Dirac. Starting from small $\lambda$ and scanning to higher values, 
the neutrinos again split,
but now the lightest neutrino has  
 4 degrees of freedom. This is enough to match  the number of bosonic ones, and the potential need not become
negative as $\lambda$ increases. However, if the lightest neutrino becomes too heavy,  at some point the number of effective degrees 
of freedom will be again dominated by the photon and the graviton. This critical mass is controlled by the initial value of the vacuum energy, i.e.
the cosmological constant $\Lambda_4$. Numerically one finds that the lightest neutrino cannot be heavier than $7.7$ meV. 
This limit is better understood by looking at Fig. \ref{dirac_cero}. 
The lines which hit the 
$\lambda=1$ limit  before reaching the blue area correspond to runaway vacua everywhere throughout the scan. Thus, there is no transition to AdS (or viceversa)
and hence the AdS distance conjecture is not violated. The limiting value corresponds to the quoted 7.7 meV. For illustrative purposes we plot the different shapes of the potential for different values of $\lambda$ for the particular case that the experimental case saturates the bound, $m_\nu^{\text{exp}} = 7.7$ meV. This is consistent with the AdS Distance Conjecture if the scanning stops at $\lambda=1$, so we never cross Minkowski space. Hence, consistency with the AdS Distance conjecture implies an upper bound on the lightest neutrino mass,  $m_\nu^{\text{exp}} \leq 7.7$ meV. Note that values $\lambda_\text{max}$ larger than 1 would only make this bound stronger.

One could think that in the case of Majorana neutrinos, the combination of the two lightest generations of neutrinos 
also provides 4 degrees of freedom, just like a Dirac fermion, so perhaps one could scape and  not  go to AdS as $\lambda$ increases.
However this would require {\it both } generations of neutrinos to be lighter than $7.7$ meV which is in contradiction with
neutrino oscillation data, both for normal and inverted hierarchy.

Note that the upper bound on the lightest Dirac neutrino mass may shed some light into the EW hierarchy problem, as we already
mentioned in the non-SUSY AdS conjecture case.
Since $m_1=Y_1|H_0|$, an upper bound on $m_1$ corresponds to an upper bound
on the Higgs vev at fixed Yukawa coupling, $|H_0|\leq m_1/Y_1$. Thus, one can argue that the Higgs vev (and hence naturally its mass) is small because otherwise
the 3D radion potential would violate the AdS distance conjecture. This explanation is predictive, since then the stability of the Higgs vev
about its observed value (and not higher) would imply a lightest neutrino mass around 7.7 meV. 
This was already remarked in \cite{IMV2, Gonzalo}.

The above results assumed normal hierarchy for the neutrinos. One obtains similar results for the IH case, the same argumentations apply. 
 Numerically, the upper bound on the lightest neutrino mass changes. The mass at which the AdS vacua starts to be generated for IH Dirac neutrinos is 2.56 meV, as already found in \cite{IMV1}. IH Majorana neutrinos would be ruled out in the same way as NH Majorana.

Let us finally emphasize that these results hold under the assumption that these homogeneous scannings are allowed by the UV quantum gravity completion of the SM. In the following, we will analyze how sensitive are the results to the specific scanning trajectory under consideration. We first consider an alternative non-homogeneous scanning and secondly, we allow for a variation of the 4d cosmologican constant on neutrino masses. 

\subsection{Non-homogeneous scanning}
\label{sec:nonhomogen}

Let us  now consider the possibility of non-homogeneous scannings, where the three neutrino masses change in a different way with $\lambda$. 
The simplest case is a scanning in the lightest neutrino Yukawa coupling as follows
\begin{align}
m_{\nu_{1}}(\lambda) & =\lambda m_{\nu_{1}}^{\text{exp}}\nonumber \\
m_{\nu_{2}}(\lambda) & =\sqrt{(m_{\nu_{1}}^{\text{exp}})^{2}+\Delta m_{21}^{2}}\nonumber \\
m_{\nu_{3}}(\lambda) & =\sqrt{(m_{\nu_{1}}^{\text{exp}})^{2}+\Delta m_{21}^{2}+\Delta m_{32}^{2}}
\end{align}
Thus, we scan on the Yukawa coupling of the lightest neutrino mass while keeping the other two Yukawas fixed. In the Majorana case, we find that, as we vary the lightest neutrino mass,
we remain all the time in the region of AdS vacua and there is no transition to Minkowski. 
Hence the AdS distance conjecture is not violated and we get no constraint on Majorana neutrino masses.
In the case of Dirac masses, we find that the trajectory traverses from positive to negative vacuum energy as $\lambda$ increases, which would be against the AdS Distance conjecture, unless one imposes an upper bound on the
lightest neutrino mass of order 7.7 meV. This ensures that the trajectory never actually crosses the flat space limit. 


Hence, we recover the same bounds for Dirac masses than in the homogeneous scanning, although there are no constraints for Majorana neutrinos.
This opens up the question of what scanning is actually realized in quantum gravity. Although this cannot be answered without further UV information, we would like to remark that the homogeneous scanning seems more natural for its simplicity and 
its possible interpretation in terms of the scanning on the Higgs vev. Scanning one generation and
not the others, as in the non-homogeneous scanning above 
may look artificial. Notice that the lightest neutrino state is the lightest eigenvalue of a non-diagonal matrix, so that it is actually quite contrived to scan only on the lightest generation, given the large value of the neutrino mixing angles. 

We could  consider different scanning trajectories in the parameter space that would correspond to different lines in the previous figures. Once we choose two points in this parameter space that should yield EFTs realized in the landscape, the results are quite insensitive to the specific chosen trajectory connecting them. However, the results are highly sensitive to the choice of these two points. Recall that one of them has to sit necessarily at $\lambda=1$ as it corresponds to the values of our universe, but there is certain freedom in choosing the other one. We have given two examples in which we have chosen this second point to yield nearly vanishing mass for the three neutrinos (homogeneous scanning) or only for the lightest one (non-homogeneous scanning). But we invite the reader to envision other scenarios and extract the results from looking at the figures. Interestingly, the results happen to follow a very concrete pattern. For Majorana, one obtains that they are either ruled out or there is no constraint on the masses. Contrary, for Dirac, any non-trivial constraint is going to imply an upper bound on the lightest neutrino given by 7.7 meV, since this is the threshold value at which the minimum takes places in Minkowski space.




\subsection{Scanning of both neutrino masses and $\Lambda_4$}
\label{sec:lambdachange}

In the above discussion we have assumed that the 4D cosmological constant $\Lambda_4$ is kept
fixed while scanning on neutrino masses. 
However, it might happen that the scanning trajectories realized in quantum gravity imply a variation of both neutrino masses and the cosmological constant. Since it is not clear what the exact dependence should be, we are going to allow for a general dependence of the form $\Lambda_4 \sim m_\nu^\alpha$ and study how this affects the results. 


More concretely, let us consider the homogeneous scanning on neutrino masses of section \ref{sec:homogeneous} but allow now for an additional dependence of the cosmological constant on $\lambda$, namely $\Lambda = \Lambda_{\text{exp}} \lambda^{\alpha}$. To start with, let us focus on the case of Dirac neutrinos, i.e. 4 degrees of freedom per generation.
Whether a minimum is formed with positive or negative vacuum energy depends on the ratio between the cosmological constant and neutrino masses. In particular, if
\beq
\frac{\Lambda_4}{m^4_{\nu_1}}\sim \lambda^{\alpha-4}\frac{\Lambda_4^{\rm exp}}{(m_{\nu_1}^{\rm exp})^4}\lesssim 1
\label{ratio}
\eeq
then the minimum occurs in AdS, while otherwise the potential remains positive. This implies that the results will highly depend on whether $\alpha<4$ or $\alpha>4$, although they are insensitive to the specific value of $\alpha$ once we focus on one of these two regimes. 

If $\alpha<4$, the results are the same than in the previous section since we can choose, without loss of generality, the case of $\alpha=0$ (fixed cosmological constant) in Fig.\,\ref{dirac_cero} to represent this set of cases. For very small $\lambda$, the ratio \eqref{ratio} is bigger than one, so that the potential takes positive values. As we increase $\lambda$, we approach the flat space limit and we will generate an AdS vacuum at $\lambda=1$ if $\Lambda_4^{\rm exp}\lesssim (m_{\nu_1}^{\rm exp})^4$. Hence, in order to avoid crossing the flat space limit at a finite value of the radius, one needs to impose an upper bound on the experimental neutrino mass given by $\Lambda_4^{\rm exp}\gtrsim (m_{\nu_1}^{\rm exp})^4$, which translates to $m_{\nu_1}^{\rm exp}\leq 7.7$ meV for the cosmological constant measured in our universe.
\begin{figure}
	\centering{}\includegraphics[scale=0.385]{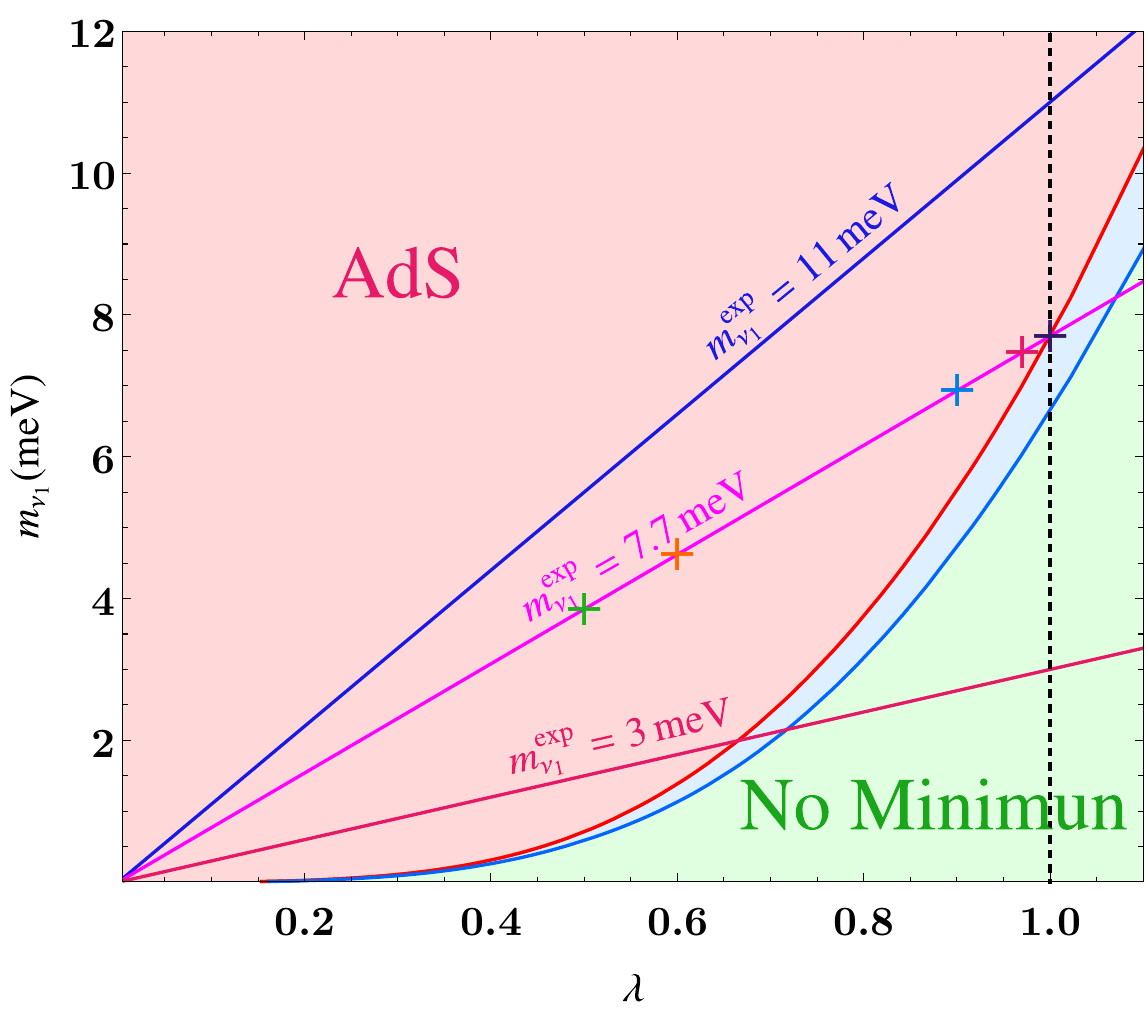} 
	\includegraphics[scale=0.3975]{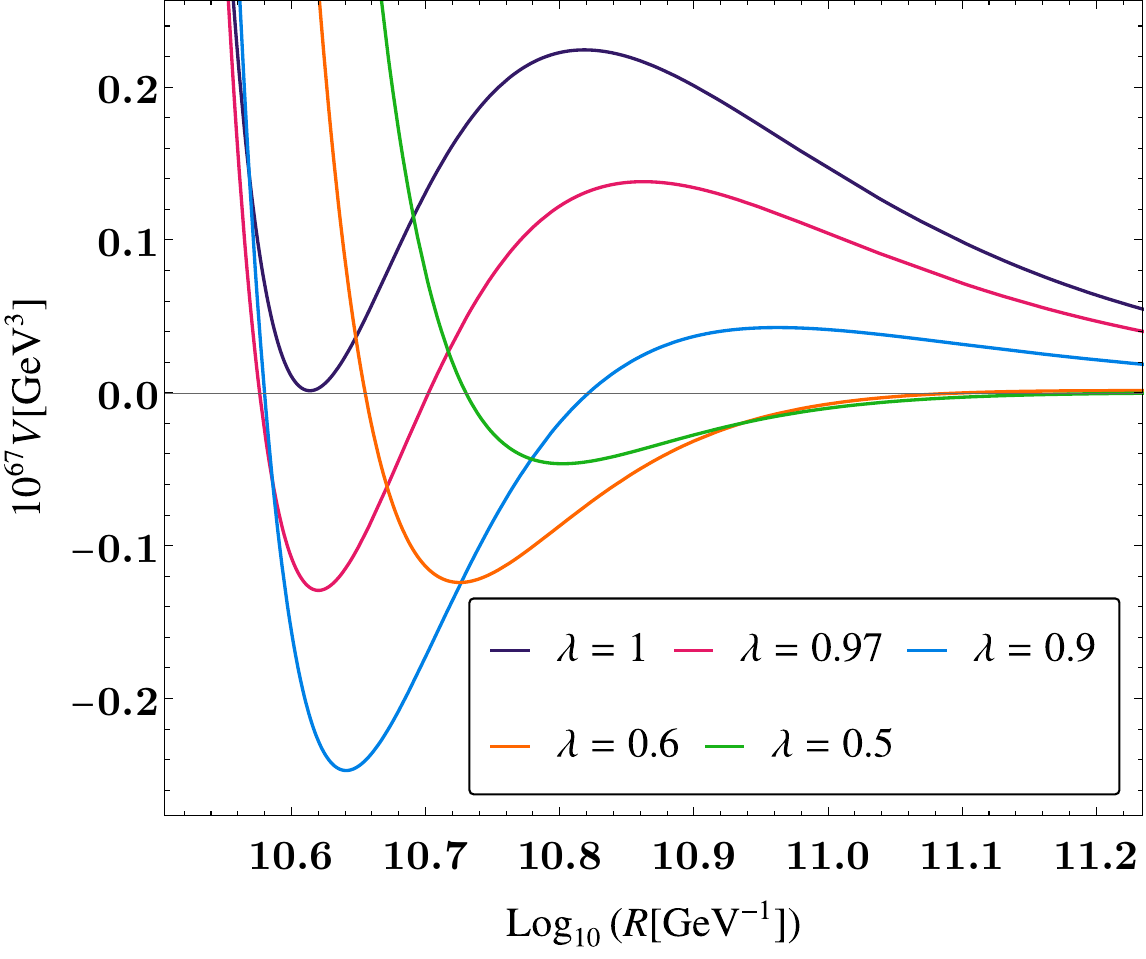}
	\caption{ Left: Lightest neutrino mass versus the scanning parameter $\lambda$ for Dirac neutrinos. The cosmological constant also changes as $\Lambda_{ 4\text{d}}(\lambda)=\Lambda_{ 4\text{d}}(\lambda=1) \lambda^{ \alpha}$, where $\alpha=10$.  The lines 
		do not cross from the AdS to the runaway area as long as the lightest neutrino mass is $ m_{\nu}^{\text{exp}} \geq 7.7$ meV. Right: In this figure we show how the potential changes along the scanning for $ m_{\nu}^{\text{exp}}$ slightly above 7.7 meV. Coloured crosses in the left hand side correspond to the coloured curves in the plot on the right hand side.}
	\label{dirac_diez}
\end{figure}

If $\alpha>4$, one gets the opposite behaviour. For very small $\lambda$, the ratio \eqref{ratio} becomes smaller than one, implying that there is a minimum in AdS. As  $\lambda$ increases, this minimum will cross the flat space limit unless one has $\Lambda_4^{\rm exp}\lesssim (m_{\nu_1}^{\rm exp})^4$, so that the ratio \eqref{ratio} remains always smaller than one throughout the scanning $0\lesssim \lambda\leq 1$. To illustrate this, we have plotted the results for $\alpha=10$ in Fig. \ref{dirac_diez}. When $\lambda$ is very small (see the green line in the  right-hand side plot) there is an AdS vacua. As we increase $\lambda$ the minimum goes deeper at first (see orange curve) but at some point (see blue curve) it stops and $V_0$ starts increasing (see pink curve) until it reaches Minkowski (see purple curve) and would become dS if the scanning continued. In  the right hand side plot of Fig. \ref{dirac_diez} we are choosing an experimental mass for the lightest neutrino equal to 7.7 meV, so that the scanning stops ($\lambda=1$)  precisely when we reach Minkowski. If the experimental mass was smaller everything would happen \textit{before}, as can be seen in the left plot. However, for masses larger than 7.7 meV (see blue curve in the left plot) everything would happen \textit{later} and we would not reach Minkowski in the scanning range $0\lesssim \lambda \leq 1$. Hence, for $\alpha> 4$, consistency with the AdS Distance conjecture implies a lower (instead of an upper) bound on the lightest neutrino mass of order 7.7 meV. This case includes the situation in which only the cosmological constant, and not the SM particle masses, scans in the landscape (so $\alpha\rightarrow \infty$). This resembles the Bousso-Polchinski \cite{BP} or KKLT \cite{KKLT} proposals. The bound we obtain corresponds then to an upper bound on the cosmological constant of the form $\Lambda_4 \lesssim m_\nu^{4}$. 

\begin{figure}
	\centering{}\includegraphics[scale=0.3505]{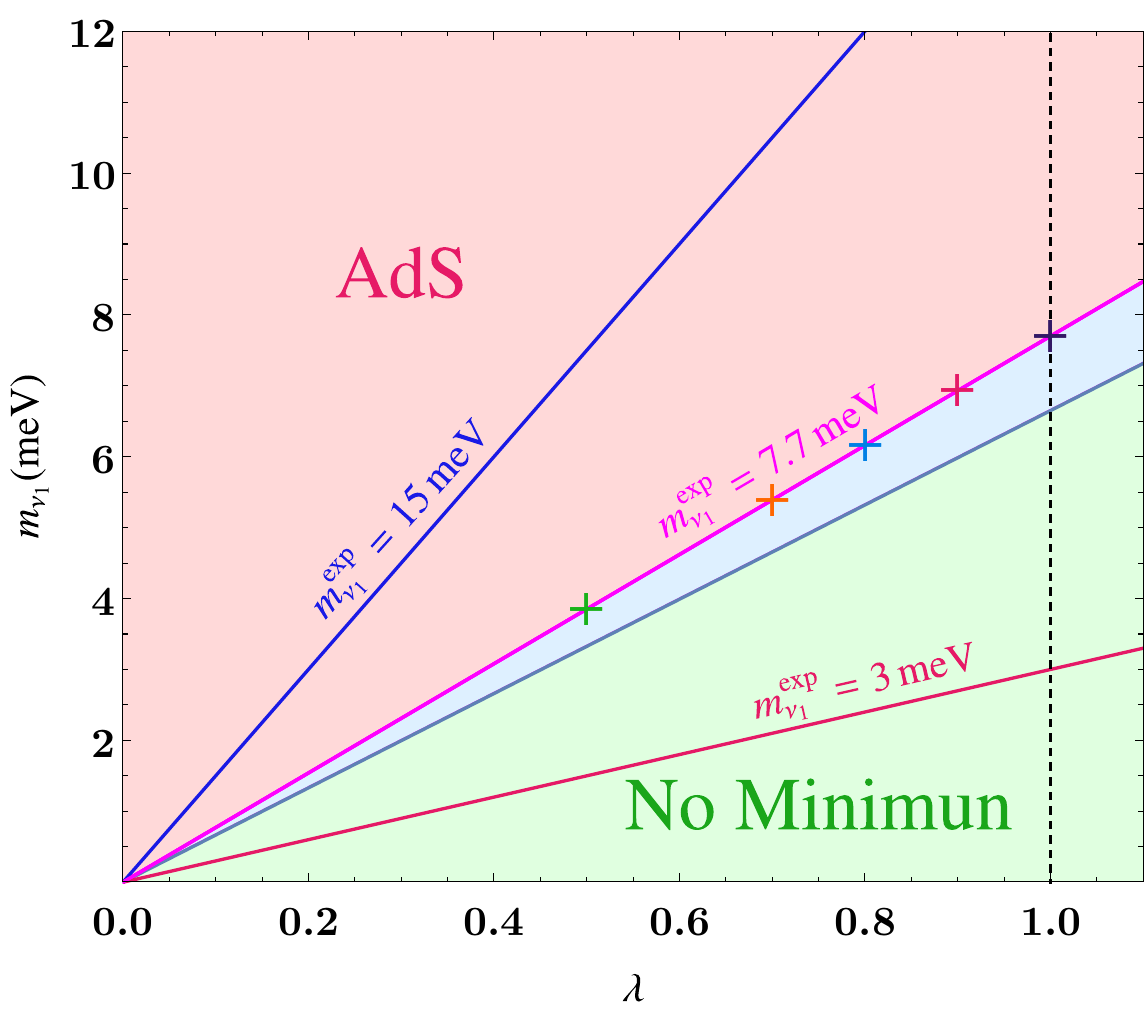} 
	\includegraphics[scale=0.37]{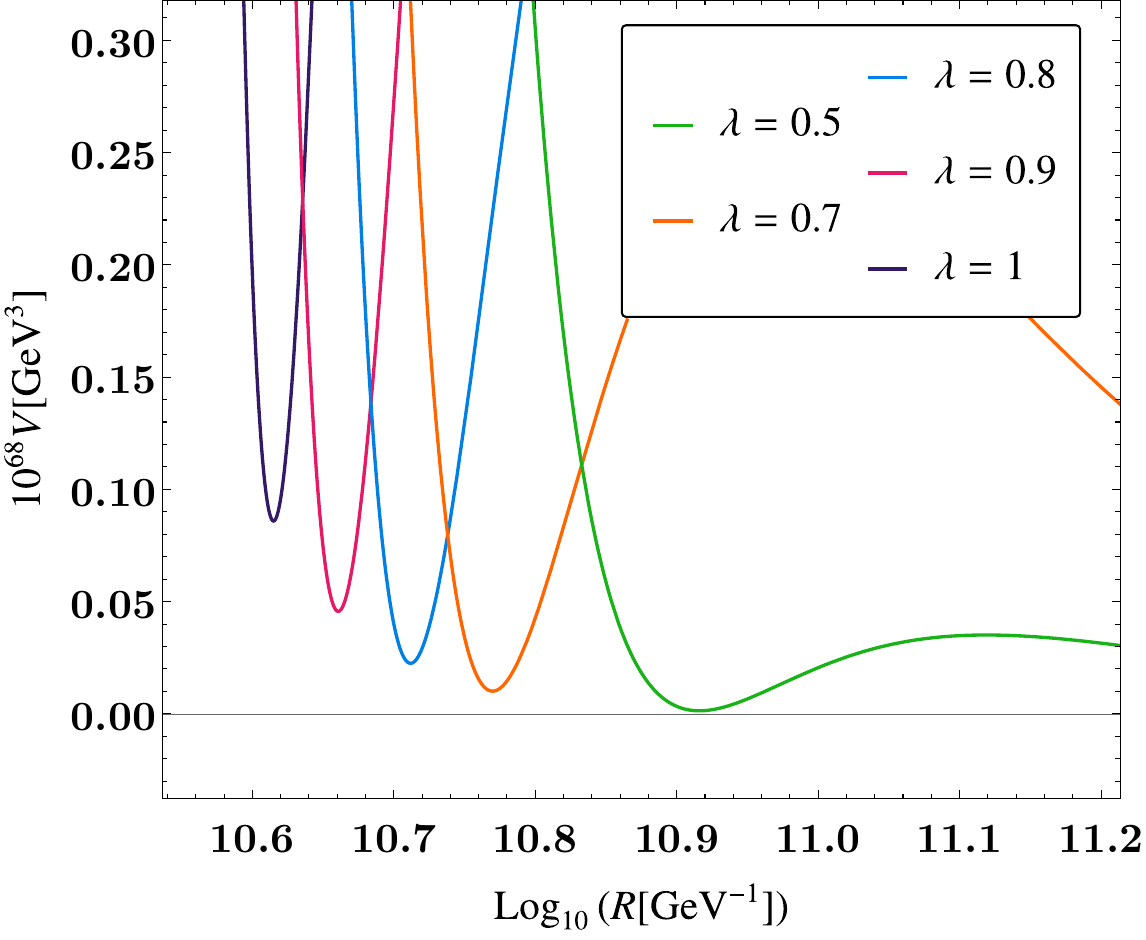}
	\caption{  Left: Lightest neutrino mass versus the scanning parameter $\lambda$ for Dirac neutrinos. The cosmological constant changes as $\Lambda_{ 4\text{d}}(\lambda)=\Lambda_{ 4\text{d}}(\lambda=1) \lambda^{ \alpha}$, where $\alpha=4$.  Straight lines never cross from the AdS to the runaway area so there is no violation of the AdC. Right: In this figure we show how the potential changes along the scanning for the particular value  $ m_{\nu}^{\text{exp}} = 7.7$ meV. We find that a tower must become light in the limit $\lambda \rightarrow 0$. This tower is the KK tower. Coloured crosses in the left hand side correspond to the coloured curves in the plot on the right hand side.}
	\label{dirac_cuatro}
\end{figure}

From the above discussion, it is clear that inconsistency with the AdS Distance conjecture occurs when one changes the tendency of the ratio \eqref{ratio}, so that it does not remain smaller (or larger) than one during the entire scanning range $0\lesssim \lambda \leq 1$. If we start with an AdS vacuum for very small $\lambda$ (as happens if $\alpha>4$), we should continue in AdS when $\lambda=1$, implying a lower bound on the experimental neutrino mass. Contrary, if we start with a positive runaway for very small $\lambda$ (as happens if $\alpha<4$), then we should continue with a positive vacuum energy at $\lambda=1$, implying instead an upper bound on the experimental neutrino mass. This also implies that the case $\alpha=4$ is automatically consistent with the AdS Distance conjecture, as the ratio \eqref{ratio} remains constant along the scan. We have illustrated the case $\alpha=4$ in Figure \ref{dirac_cuatro}. The scanning trajectories correspond to straight lines in the left-hand side plot that never cross the flat space limit. Thus, if we start for instance in $AdS$ for small $\lambda$, we will remain in the $AdS$ region at $\lambda=1$. Notice that, for this case, even if in the limit $\lambda \rightarrow 0$ the family of vacua approaches the flat space limit $V_0\rightarrow 0$, this occurs for a radius $R \rightarrow \infty$, so the KK tower becomes light in agreement with the AdC. Finally, even though the case $\alpha=4$ is automatically consistent with the AdC and it does not yield any bound on neutrino masses, notice that it is already implying by definition a correlation of the form $\Lambda_4\sim m_\nu^4$, as suggested experimentally.

 To sum up, unlike in the previous sections, the results here highly depend on what scanning trajectories may be realized in quantum gravity. In particular, depending on whether $\alpha<4$ or $\alpha>4$, we get an upper or lower bound on the lightest neutrino mass set by the threshold value of 7.7. meV.
 Interestingly, the Non-SUSY AdS conjecture favours those with  $\alpha \leq 4$ so that an AdS vacuum is never formed. In such a case, both the Non-SUSY AdS conjecture and the AdS Distance conjecture imply the same upper bound on neutrino masses.

The results presented so far apply to Dirac neutrinos with Normal Hierarchy, but we could also consider the cases of IH Dirac neutrinos and Majorana neutrinos. On the one hand, for IH Dirac neutrinos we obtain similar results as for NH Dirac neutrinos: an upper (lower) bound of 2.56 meV  on the mass of the lightest neutrino for scannings with $\alpha < 4$ ($\alpha > 4$). On the other hand, Majorana neutrinos are ruled out by scannings with $\alpha<4$. 
For $\alpha>4$, the potential exhibits an AdS vacuum in the limit $\lambda \rightarrow 0 $,  whose depth goes to 0 as the KK tower comes down, so we find agreement with the AdC, just like we found for Dirac neutrinos. In the limit $\lambda \rightarrow 1$ the vacua is still in AdS so there is no constraint on the experimental mass of the lightest neutrino. To sum up, Majorana neutrinos are ruled out by $\alpha<4$ scannings and left unconstrained by $\alpha=4$ and $\alpha>4$  scannings. 

\subsection{Festina lente, neutrinos and electrons}
We would like to finish this section by making some comments on the application of the {\it Festina Lente} (FL) constraints recently considered  in \cite{festina,festinanew}  and their combination with 
the above ideas. According to the FL hypothesis, charged particles under a U(1) gauge field and with mass $m_e$ in a dS vacuum should verify the constraint
\begin{equation} m_{e}^4\ \gtrsim \ 2g_{\text{em}}^2\rho_{\Lambda} \ .
\label{FL0}
\end{equation}
Note that any charged particle of the SM, and in particular the electron, verify this equation. Still, it is a very non-trivial upper bound on the c.c.,
whose natural value is naively $M_p^4$.  Interestingly, this is an upper constraint on the c.c. whereas we have seen that the AdS conjectures give us 
a lower bound $\Lambda_4 \gtrsim m_{\nu}^4$, so we can write for the electron and the lightest neutrino masses
\beq
m_e^2 \geq \ g\ \Lambda_4^{1/2}\ \gtrsim \ g\ m_{\nu}^2.
\label{FLcon}
\eeq
Note that this equation implies an upper bound on the gauge coupling
\beq
g \ \lesssim \frac {m_e^2}{m_{\nu}^2} \ .
\eeq
Again,  this is trivially satisfied  in the SM, but it also tell us that a limit $m_e/m_\nu\rightarrow 0$ in a dS background would be in the Swampland or else 
an infinite  tower of charged particles should appear in that limit. 
Concerning the EW hierarchy problem, with fixed Yukawa couplings $Y_e$, $Y_\nu$, combining FL and the AdS conjectures one
obtains upper and lower bounds on the Higgs vev
\beq
\frac {g}{Y_e^2}\ \Lambda^{1/2} \ \lesssim \ \left| \langle h_0 \rangle \right| ^2   \  \lesssim\ \frac {\Lambda^{1/2}}{Y_\nu^2} \ .
\eeq
The range is very wide, but it is interesting the complementary role of the FL and AdS Swampland constraints and the fact that
there is no inconsistency among them.

Let us go back to the AdS distance conjecture, and consider a homogeneous scanning of fermion masses corresponding to
a Higgs scanning.  One can write $m_e,m_\nu \rightarrow \lambda m_e, \lambda m_\nu $ and $\Lambda \rightarrow \lambda^\alpha \Lambda $.
Assuming this scanning exist in the landscape, Eq. (\ref{FLcon}) would imply that  one is forced to have $\alpha = 4$, which corresponds to an scaling of the form $\Lambda_4\sim m_\nu ^4$. As we showed, 
this value is critical in that the AdS distance conjecture is always satisfied, leading to no further constraint on neutrino masses from the
AdS conjecture. 
However that would be too fast a conclusion since  having a minimum value 
$\lambda_\text{min}=\Lambda^{1/4}/m_e \simeq 10^{-9}$ for the scanning would be enough to get no contradiction with FL even if $\alpha\neq 4$. 
In fact such non-vanishing value for $\lambda_\text{min}$ was shown to be needed in homogeneous scannings in order to avoid other 
AdS vacua to appear close to the QCD scale \cite{Gonzalo}.  Thus, the constraints from FL are not strong enough to restrict further our discussion 
about neutrino masses although the fact that FL and AdS constraints are complementary is rather intriguing.

\section{Quintessence, the dS/AdS conjectures and neutrino masses}

In the previous sections, we have always assumed that we live in a 4d dS vacuum. However, it could also be that the accelerated expansion of our universe is not described by a cosmological constant but by a quintessence rolling scalar field. This latter possibility is in fact motivated by the de Sitter conjectures, at least near the asymptotic boundaries of the moduli space. For completeness, in this section we consider this second scenario in which we start already with a quintessence-like potential in 4d. Upon circle compactification, we will still have a runaway potential for the quintessence field, so we cannot use anymore the AdS Swampland conjectures to provide non-trivial bounds as they only apply to actual minima. But we can check the constraints coming from the dS conjecture applied to lower dimensional compactifcations of the SM. Interestingly, even if the $dS$ conjecture is satisfied in 4d, quantum corrections can violate it in 3d unless, again, we have enough light fermionic degrees of freedom. Therefore, we recover similar bounds on neutrino masses as in the previous sections from applying yet another different Swampland conjecture.

Let us briefly review how single field models of quintessence, with canonical kinetic terms and a potential $U(\phi)$ are constrained by cosmological data
(see e.g.\cite{tsujikawa,linder}).
 Using the background equations one can obtain expressions for the derivatives of the potential in terms of the derivative of the equation of state $w$ with respect to the scale factor. One finds that the experimental observation $\frac{dw}{da} \lesssim  1$ implies:
\begin{equation}
\epsilon_{V}=\frac{c_{V}^2}{2}\equiv \frac{1}{2} (\frac{U'(\phi)}{U(\phi)})^2(M_{p}^{4d})^2 \lesssim 1
\end{equation} 
and 
\begin{equation}
m_{\phi}^2 =\frac{\partial^{2}U}{\partial\phi^{2}} \sim \sqrt{ \frac{\rho_\phi^{0}}{M_{P}^2}}\sim 10^{-42} \text{GeV}.
\label{experimental_epsilon}
\end{equation}
We have introduced a subindex in $c_{V}$ to distinguish it from the parameter $c$ that appears in the statement of the $dS$ conjecture Eq. \eqref{dsone}.

It is well known that the $dS$ conjecture Eq. \eqref{dsone} is in tension with observations from Inflation and Dark Energy \cite{dscosmo1,dscosmo2,dscosmo3}. This tension is translated into bounds on the parameter $c$. The strongest bounds come from single-field inflationary models, which give $c\lesssim 0.01$ \cite{dscosmo2}. Regarding single-field quintessence models a sharp bound $c<0.6$ (or $\epsilon<0.18$) was derived in \cite{dscosmo1} imposing consistency of the $dS$ conjecture and experimental observations.  In this work, we are concerned only with the implications of the Swampland program on neutrinos, so we will assume that our Universe is described by a simple single-field quintessence model that satisfies the $dSC$ and is in agreement with observations, so $c< 0.6$. We expect similar results from more elaborate quintessence models.
 

\subsection{Constraints on the Casimir potential}
The dimensional reduction of the SM+Quintessence in a circle gives:
\begin{align}
\Gamma & =\int d^{3}x\sqrt{-g_{3}}\biggl\{2\pi r\left(M_{p}^{4d}\right)^{2}\frac{1}{2}\mathcal{R}^{3d}+2\pi r(M_{p}^{4d})^{2}\frac{\partial^{i}R\partial_{i}R}{R^{2}}-2\pi r\left(\frac{r}{R}\right)^{2}U(\phi)\\
& -V_{1L}(R)+2\pi r\sum_{n=1}^{\infty}\frac{1}{2}\left(\partial^{i}\phi_{n}\partial_{i}\phi_{n}-m_{n}^{2}\phi_{n}^{2}\right)\biggr\}
\label{reduction_quintessence}
\end{align}
where $m_n^2= (r/R)^{2}( m_{\phi}^2 + n^2/R^2)$. The quintessence field is very light, $m_\phi \sim 10^{-42}$ GeV, when compared to the neutrinos and the cosmological constant, $m_\nu \sim \rho_{\phi}^{1/4} \sim 10^{-12}$ GeV. For this reason its contribution to the Casimir energy $V_{1L}$ is essentially that of a massless particle. Previously, we had four massless bosonic degrees of freedom (graviton + photon) and now we simply need to add one more, so we have five in total. Hence, the one-loop term is therefore not very sensitive to the details of the quintessence model as long as it involves only one scalar. 

We would like now to study whether the effective 3D potential is consistent with the dS conjecture as applied to AdS, as proposed in \cite{lpv} (see also \cite{heliudson}) and discussed in the introduction. More concretely, we will assume that it is satisfied in 4d and check the constraints that appear from imposing that it is also satisfied in 3d, where it reads
\begin{equation}
(M_{p}^{3d})^{1/2}\frac{|\nabla V|}{|V|}>c.
\label{dsuno_3d}
\end{equation}
Notice the derivatives in the last formula are with respect to the 3d field $\phi’ = r^\frac{1}{2} \phi$.

In addition to the radion field $R$, we have an additional scalar $\phi$ coming from the 4d quintessence field, so \eqref{dsuno_3d} becomes
\begin{equation}
\label{dS3}
\frac{1}{|V|}\sqrt{\left(R\left|\frac{\partial V}{\partial R}\right|\right)^{2}+\left(M_{p}^{4d}\left|\frac{\partial V}{\partial\phi}\right|\right)^{2}}>c.
\end{equation}
Before showing the numerical results, let us identify the regions in field space that could potentially yield a violation of this bound. For very large radius, the potential is dominated by  the tree level contribution $V_{\text{tree}}=2\pi r\left(\frac{r}{R}\right)^{2}U(\phi)$. Replacing this into \eqref{dS3}, we get
\begin{equation}    
\sqrt{4+c_{V}^2}>c.
\end{equation}
Consistency of the de Sitter conjecture in 4d implies  $c_V \geq c$, which guarantees that the above bound is always satisfied. Hence, the runaway region of the potential at large radius automatically satisfies the conjecture if it was already satisfied in 4d.

As we decrease the radius, the quantum corrections to the 3d potential become more important and need to be taken into account. In fact, the bound \eqref{dS3} risks to get violated at the critical points where $\partial_R V=0$. At those points, the first term in \eqref{dS3} vanishes and the bound becomes
  \begin{equation}
  \frac{c_{V}}{|1+\frac{V_{1L}}{V_{\text{tree}}}|} \ >\ c\ ,
  \label{slowroll}
  \end{equation}
  where we have used the fact that only the tree level potential depends on the quintessence field.
  The above condition  can be violated
   if the one loop contribution becomes comparable or larger than the tree level one. Interestingly, this occurs precisely when an AdS vacuum is formed. In fact, the larger the neutrino masses are, the deeper the AdS vacuum is and the larger the violation of \eqref{slowroll}. Hence, consistency with the conjecture in 3d will again imply an upper bound on neutrino masses, as we compute in detail below.

Let us emphasize that we are analyzing the version of the dS conjecture proposed in \cite{lpv,heliudson} that contains the absolute value for the potential in the denominator in \eqref{dsuno_3d}. This is specially motivated in AdS space \cite{lpv}.  Without this absolute value (as in the original dS conjecture \cite{dS3}), the bounds on neutrino masses disappear as the conjecture gets automatically satisfied for AdS minima. It was also proposed in \cite{dS3,Garg:2018reu} that the theory could still be consistent with quantum gravity even if this bound on the first derivative of $V$ is violated as long as the second derivative satisfies $V''\leq -c'V$. However the status of  this ``refinement" is weak,  since at the moment it is not clear what the condition on the second derivative must be in AdS. 
We will thus restrict ourselves to the condition in Eq. (\ref{dsuno_3d}).



%
%
  \subsection{Bounds on neutrino masses}
     \begin{figure}
  	\centering{}\includegraphics[scale=0.40]{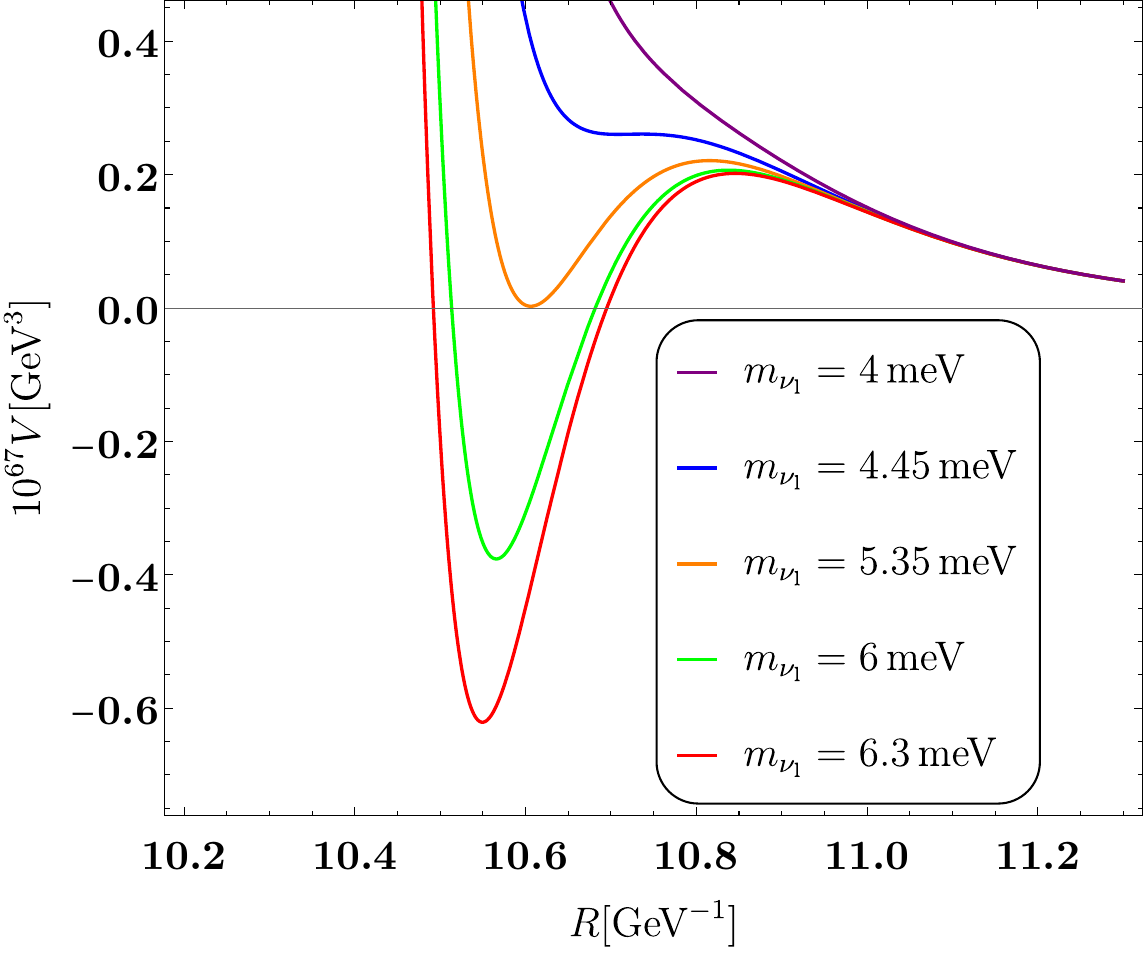} \
  	\caption{  The radion potential for various values of the lightest (Dirac) neutrino mass with fixed quintessence mass. }
  	\label{dSPot}
  \end{figure}

Let us first consider the case of NH Dirac neutrinos.
  The SM+quintessence potential for NH Dirac neutrinos is plotted in Fig. \ref{dSPot} for several values for the lightest neutrino mass. We see that there are $AdS$ vacua if the lightest neutrino is heavier than 5.35 meV. This differs from the results in section \ref{introneu} due to the additional contribution of the quintessence field to the Casimir potential, which increases the number of bosonic degrees of freedom as explained above.
  
  
\begin{figure}
	\centering{} \includegraphics[scale=0.4]{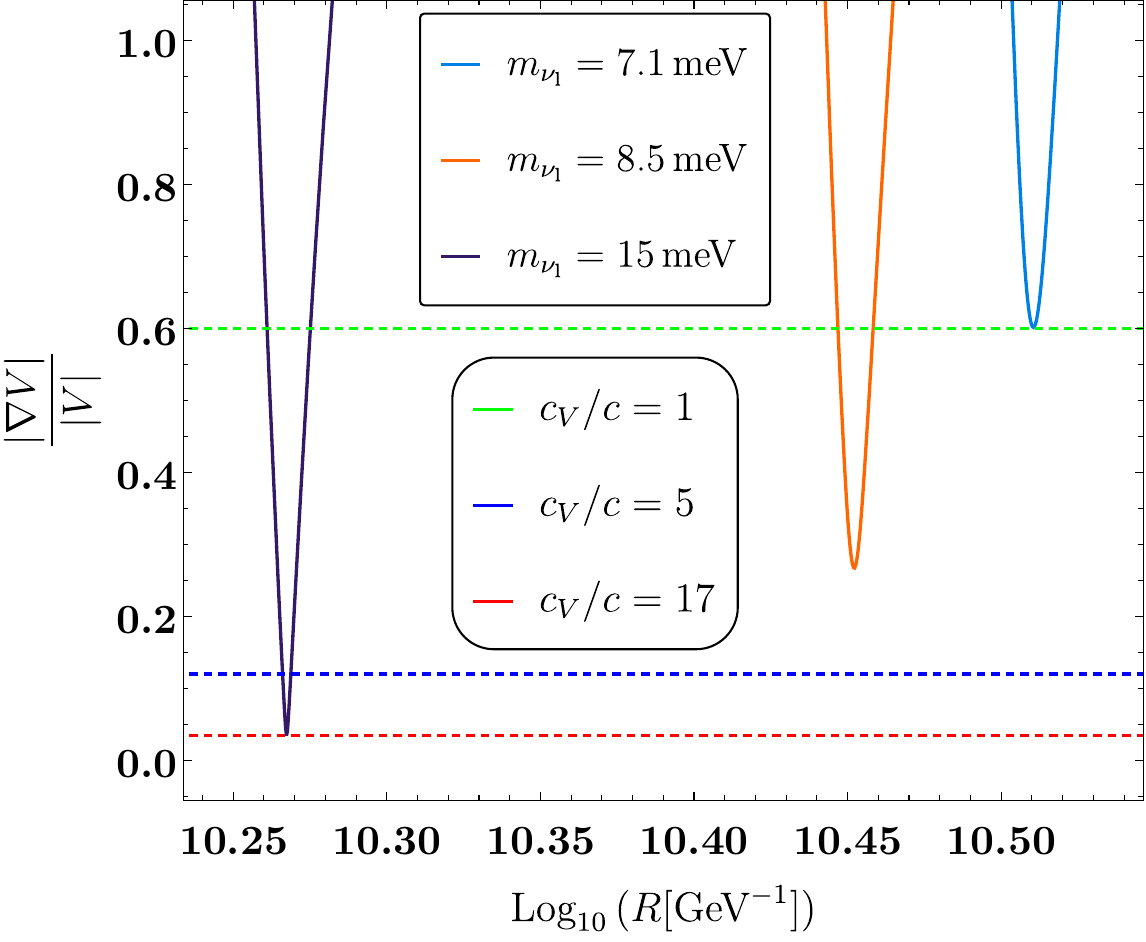}\
	\caption{  $(M_{p}^{3d})^{1/2}\frac{|\nabla V|}{|V|}>1$ for a quintessence potential with $c_{V}=0.6$ for Dirac neutrinos in NH, with  $c=0.6,0.12  \, \text{and} \, 0.035$. For $c=0.6$ we find the bound $m_{\nu_{l}} \leq 7.1$ meV. However, for $c=0.035$ we find the weaker bound $m_{\nu_{l}} \leq 15$ meV. }
	\label{bounds}
\end{figure}
As mentioned, the $dS$ conjecture in Eq.\eqref{dsuno_3d} is violated in a deep enough AdS vacuum. Recall that the depth of the vacuum is determined by how heavy the lightest neutrino is, as can be seen in Figure \ref{dSPot}. Notice that the bigger the ratio $\frac{c_{V}}{c}$ the deeper the vacua can be without violating the conjecture. This is further represented in Fig. \ref{bounds} where we plot $\frac{|\nabla V|}{|V|}$ as a function of the radius for different values of the lightest neutrino mass. 
The larger the neutrino masses, the more difficult to satisfy the dS bound since $\frac{|\nabla V|}{|V|}$ gets smaller. The specific bound on neutrino masses depends on the value of $c_V/c$. Of course, the larger this ratio is, the weakest the bounds on neutrino masses become. However, experiments seem to already require $c_{V}\lesssim 0.6$, so a very large ratio $c_V/c$ is in tension with having $c\sim \mathcal{O}(1)$ as proposed in the dS conjecture. To give some examples, if we set $c_V=0.6$, we obtain that the lightest neutrino should satisfy $m_{\nu_{l}} \leq 7.1$ meV for $c=0.6$, or  $m_{\nu_{l}} \leq 15$ meV for $c=0.06$. Hence, the bound does not crucially depend on the precise coefficient $c$ of the conjecture, giving always the same order of magnitude.

\begin{figure}
	\centering{}  \includegraphics[scale=0.45]{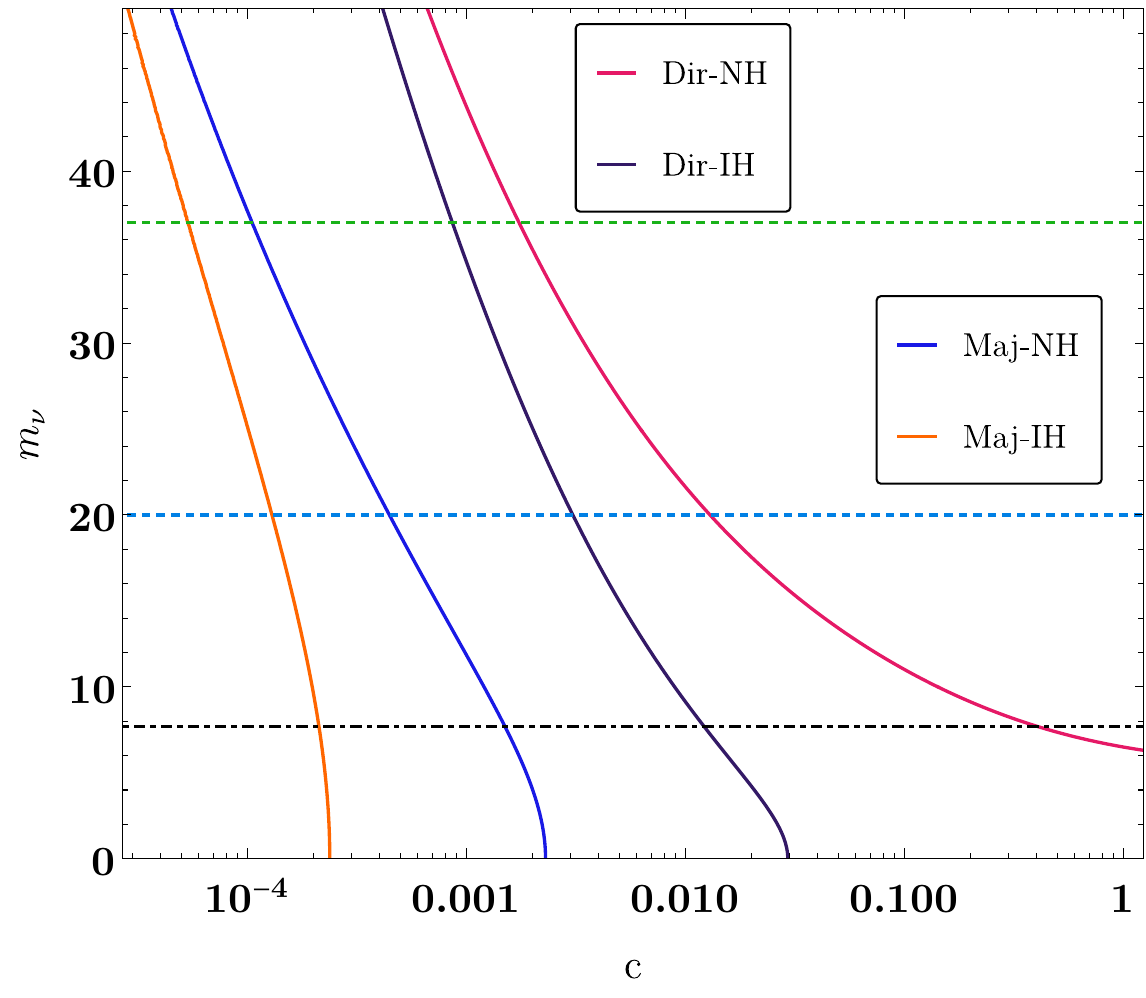} \
	\caption{Upper bound on the mass of the lightest neutrino provided by the dS conjecture as applied to AdS vacua. The conjecture requires 
		$c\sim {\cal O}(1)$, so that only Dirac fermions with normal hierarchy are consistent, while Dirac neutrinos with IH or Majorana neutrinos 
		require $c\lesssim 0.001$. 
		For reference, we plot the two upper horizontal lines which correspond to present  cosmological limits  on the
		lightest neutrino mass for NH,  the green line from \cite{GAMBIT} and  the light blue line from \cite{olga}. The lowest horizontal line, in black, is the upper limit
		from the AdS conjectures in previous sections.}
	\label{fig:gambit}
\end{figure}

In Fig.  \ref{fig:gambit}  we plot the upper bound on the lightest neutrino mass for different values of $c$.
 We see that a value $c=0.1\sim1.0$ as expected from the
dS conjecture may only be reached  for Dirac fermions in normal hierarchy NH.  This happens if $m_{\nu_1}\lesssim  10$ meV, very similar to the bounds obtained from the AdS conjectures with a cosmological constant. Dirac masses with IH would rather require $c\lesssim 0.01$ and Majorana neutrinos $c\lesssim 0.001$, which are values hardly consistent with the dS conjecture as applied to the 3D AdS vacua. Thus these Swampland arguments strongly favour the case of Dirac fermions with normal hierarchy over the other options.

It is interesting at this point to review the status of the experimental constraints on the mass of the lightest neutrino. The best limits come from cosmological
bounds, CMB temperature fluctuations and polarization,  BAO observations and Supernova   Ia  data. Such data restrict the sum of the neutrino masses.
Thus e.g. ref.\cite{GAMBIT} yields $\sum_i m_{\nu_i} < 0.139(0.174)$ at 95\% CL  for NH(IH) neutrinos, corresponding to bounds for the lightest  neutrino  $m_{\nu_1}<37(42)$ meV for
NH(IH). The more recent ref. \cite{olga}  finds the stronger constraint $\sum_im_{\nu_i}<0.09$ at 95\% CL.  This result slightly disfavours inverted hierarchy and yields
$m_{\nu_1}<20$ meV for normal hierarchy. This value is only around a factor 3 larger than the upper bound of 7.7 meV we derived using the AdS conjectures in
previous sections. We show these experimental limits in Fig. \ref{fig:gambit} as well as the 7.7 meV bound for comparison with the results for the Quintessence case.
Although the cosmological limits asume the standard $\Lambda$-CDM scenario, similar results are expected for quintessence models with $w\simeq -1$ \cite{hannestad}.
Limits on the sum of the three neutrino masses are expected to be improved substantially in the next decade from data from DESI \cite{DESI} and Euclide \cite{euclid}.
Limits on the lightest neutrino mass from $\beta$-decay o neutrino-less double-$\beta$ decay are much weaker. The latter process is only sensitive to lepton-number
violating neutrino mass, like in the case of pure Majorana neutrinos. 
In this connection, note  that by Dirac neutrino  we mean in this paper  a neutrino with four degrees of freedom almost degenerate in mass. This allows for lepton 
number violation couplings and hence such ``Dirac" neutrinos could in principle be detected in  $\nu$-less double $\beta$-decay experiments. However 
the reach of these experiments for neutrino masses is much above the constraints coming from cosmology.

It is remarkable that the $dS$ conjecture as applied to the SM 3D AdS vacua  implies similar bounds on neutrinos as the other two $AdS$ Swampland conjectures we have studied in previous sections. Regardless of the specific value of the coefficient $c$, we always recover an upper bound on the Dirac neutrino mass of order meV, since it is correlated to the value of the cosmological constant. It is indeed tempting to conclude that 
 the numerical coincidence between neutrino masses and the cosmological constant observed in our universe is a reflection of a deep quantum gravity structure.

         

 \section{Discussion}
 
 The fact  that the scale of neutrino masses is quite similar to the cosmological constant scale is quite intriguing. While it could be just a coincidence, it may be telling us something important about possible connections between Particle Physics, Cosmology and Quantum Gravity. 

Recently such a possible connection has been found within the context of  the Swampland program in Quantum Gravity and String Theory. It has been shown
that  the application of the non-SUSY AdS conjecture \cite{Ooguri:2016pdq} to the circle compactification of the SM  gives constraints on the mass of the lightest neutrino
$m_{\nu_1}\lesssim 7.7$ meV \cite{IMV1}. This arises essentially from the condition that 3D AdS SM vacua do not form, which roughly requires $m_{\nu_1}\lesssim \Lambda_4^{1/4}$.
The lightest neutrino must also have four degrees of freedom,  which occurs for Dirac (or quasi-Dirac) neutrinos, so that Majorana neutrinos are not allowed.
 This gives a very elegant connection between the scales of neutrino masses and the cosmological constant. In fact it also tells us that  Minkowski non-SUSY vacua are inconsistent if the lightest neutrino is not massless, so an accelerated universe is predicted, in agreement with observations.
While these results are very encouraging, and the non-SUSY AdS conjecture has been shown to be correct in  all top-down string theory examples up to now, it rests on the 
full stability of the 3D SM radion potential. And unfortunately,  it is not possible to guarantee full stability of the 3D  background without a more complete knowledge of the UV physics.

In this paper, we consider other Swampland conditions applied to the same class of 3D SM vacua in order to check whether similar constraints on neutrino masses
may be found without assuming stability of the 3D vacua. In particular, 
using the AdS distance conjecture  \cite{lpv} and scanning on different values for neutrino maases, one can derive constraints on the lightest neutrino mass which turn out to be numerically quite similar to the ones found 
from the non-SUSY AdS conjecture mentioned above. The neutrinos again must be Dirac.  An important difference, though,
 is that now the condition is local, and it is independent on the stability of
the 3D SM vacuum. However, it relies instead on the assumption that there is a family of vacua with different values of neutrino masses generated e.g. by scanning on the value of the Higgs vev. This assumption of scanning on the Higgs vev has already been taking in the literature in different contexts \cite{Herraez:2016dxn,Giudice:2019iwl,Kaloper:2019xfj}. We have also studied the case in which both the neutrino mass and the cosmological constant scan in a landscape, with 
$m_{\nu_1}\rightarrow \lambda m_{\nu_1}$ and $\Lambda_4\rightarrow \lambda^\alpha \Lambda_4$. For all possible scannings with $\alpha< 4$ one obtains the same bound 
with $m_{\nu_1}\lesssim 7.7$ meV upon replacing the experimental value of $\Lambda_4$. For scannings with $\alpha >4$ the bound becomes reversed, with $m_{\nu_1}\gtrsim 7.7$ meV. From a string theory perspective, these two possible regimes correspond to local vs global scannings in which we vary more significantly the SM particle masses or the 4D cosmological constant, respectively. It is interesting to remark that in the latter case, which resembles a Bousso-Polchinski scenario \cite{BP}, we obtain an upper bound for the 4D cosmological constant in terms of neutrino masses, in agreement with the tiny value observed in our universe. This is another example in which the logic of naturalness in field theory fails upon taking into account quantum gravity considerations.

It is possible that the observed accelerated expansion of the universe could be due to a runaway potential of a scalar field (Quintessence), rather than 
to the existence of a cosmological constant. Such a possibility has also been suggested in the context of the Swampland program in which the dS \cite{dS1,Krishnan} and TCC \cite{TCC} conjectures 
imply that only very short lived dS vacua, if at all, may be consistent with Quantum Gravity. If that was the case,  it is important to see whether the connection 
between neutrino masses and the cosmological acceleration still remains,  and whether constraints on neutrino masses can again be obtained.
Using a simple single scalar Quintessence extension of the SM we  explore the behavior of the radion-Quintessence scalar potential after circle compactification.
In such a system no  3D AdS stable vacua are generated, so we cannot make use of the two AdS conjectures used in the rest of the paper. 
However we find that,  applying the dS conjecture as applied to AdS vacua \cite{lpv}, again constraints on the lightest neutrino mass are found, with 
$m_{\nu_1}\lesssim 10$ meV. 
We find that only Dirac neutrinos in normal hierarchy are consistent without requiring a severe fine-tuning of the dS conjecture c  parameter, while inverted hierarchy or Majorana neutrinos 
require $c\lesssim 0.01$.

It is remarkable how three different Swampland conditions point to a connection between the mass of the lightest neutrino and the cosmological acceleration.
At present the best  experimental  results for the mass of the lightest neutrino come from analysis of cosmological constraints combined with terrestrial data 
(see e.g. \cite{GAMBIT,olga,hannestad}). For normal hierarchy e.g. ref. \cite{olga} implies $m_{\nu_1}<20$ meV, only a factor 3 above the bounds obtained
in this paper.  Thus e.g, if evidence is found for a lightest neutrino mass around 20 meV, 
the bound $m_{\nu_1}\leq 7.7 $ meV  (which applies to a c.c.) would be violated and hence, either the conjecture is wrong or a Quintessence scenario with $c\sim 0.01$ 
would be favoured (see Fig. \ref{fig:gambit}) or very light exotic fermionic Beyond SM (BSM) would be required. This latter possibility originates on the fact that we are considering only the SM particle spectrum in our analysis. However, as noticed in \cite{IMV1}, these neutrino bounds can be relaxed if there are additional light fermionic degrees of freedom (of order the neutrino scale or lighter). It would be interesting to further explore the possible implications of this scenario for dark matter. 

It is quite exciting how ideas based on Quantum Gravity and black-hole physics may turn out to lead to an explanation of the 
neutrino mass-cosmological constant coincidence, or to the prediction of new light BSM physics, and that these same ideas may be falsified by forthcoming data!

\vspace{2.0cm}

\centerline{\bf \large Acknowledgments}

\bigskip

\noindent We thank  A. Herr\'aez, J. Lesgourges, O. Mena, M. Montero and  T. Rudelius for useful discussions. 
This work is  is supported  by  the  Spanish  Research  Agency  (Agencia  Espa\~nola  de  Investigaci\'on) through  the  grants  IFT  Centro  de  Excelencia  Severo  Ochoa  SEV-2016-0597, the grant GC2018-095976-B-C21 from MCIU/AEI/FEDER, UE and the grant PA2016-78645-P. 
   E.G. is supported by the Spanish FPU Grant No. FPU16/03985. The research of IV was supported by a
grant from the Simons Foundation (602883, CV).

\bibliographystyle{jhep}
\bibliography{edubib}

\end{document}